\documentclass[10pt,aps,prl,raggedbottom,longbibliography,nobalancelastpage,reprint,citeautoscript,letterpaper,superscriptaddress]{revtex4-2}

\usepackage[compat=1.1.0]{tikz-feynhand}
\usepackage{amsmath,amssymb,bm,braket,url,dsfont}
\usepackage{mathtools}
\usepackage{graphicx}
\usepackage{xcolor}
\usepackage{longtable}
\usepackage{multirow}
\usepackage{array}
\usepackage{booktabs}
\usepackage{physics}
\usepackage{here}
\usepackage[version=3]{mhchem}
\usepackage{siunitx}

\usepackage[
draft=false,
bookmarks=true,
bookmarksnumbered=false,
bookmarksopen=false,
bookmarksopenlevel=4,
colorlinks=true,
anchorcolor=black,
citecolor=blue,
filecolor=magenta,
linkcolor=blue,
linkbordercolor={1 0 0},
urlcolor=blue,
pdfborder={0 0 1},
pdftitle={},
pdfauthor={},
pdfsubject={},
pdfkeywords={},
pdfpagemode=UseThumbs,
]{hyperref}

\newcommand{\expecval}[1]{\left \langle {#1} \right \rangle}

\newcommand{\figref}[1]{FIG.~\ref{#1}}

\newcommand{\Eqref}[1]{Eq.~\eqref{#1}}

\begin{document}

\title{
        Dirac charge in antiferromagnetic topological semimetals
}
\author{Kohei Hattori}
\email{hattori-kohei053@g.ecc.u-tokyo.ac.jp}
\affiliation{Department of Applied Physics, The University of Tokyo, {Bunkyo}, Tokyo 113-8656, Japan}

\author{Hikaru Watanabe} 
\email{hikaru-watanabe@g.ecc.u-tokyo.ac.jp}
\affiliation{Department of Physics, University of Tokyo, {Hongo}, Tokyo 113-0033, Japan}

\author{Ryotaro Arita} 
\affiliation{Department of Physics, University of Tokyo, {Hongo}, Tokyo 113-0033, Japan}
\affiliation{Center for Emergent Matter Science, RIKEN, {Wako}, Saitama 351-0198, Japan}

\begin{abstract}
Topological node of electronic bands can carry emergent charge degree of freedom such as the Berry curvature monopole of the Weyl semimetals, which results in intriguing transport and optical phenomena.
In this study, we discuss the existence of the hidden "Dirac charge" and its detection via the photocurrent response in antiferromagnetic (AFM) Dirac semimetals.
In light of the Berry curvature defined in the spin and spin-charge-mixed parameter space, we identify Dirac charges as sources or sinks of the Berry curvature in the generalized parameter space.
We demonstrate that this Dirac charge can be detected via the photocurrent driven by the spin-charge-coupled motive force.
By using real-time simulation, we find that the Dirac charge plays a significant role in the photocurrent generation in AFM Dirac semimetals.
This work reveals the hidden property of the Dirac points in AFM Dirac semimetals.

\end{abstract}

\maketitle
\begin{figure}
        \centering
        \includegraphics[width=\linewidth]{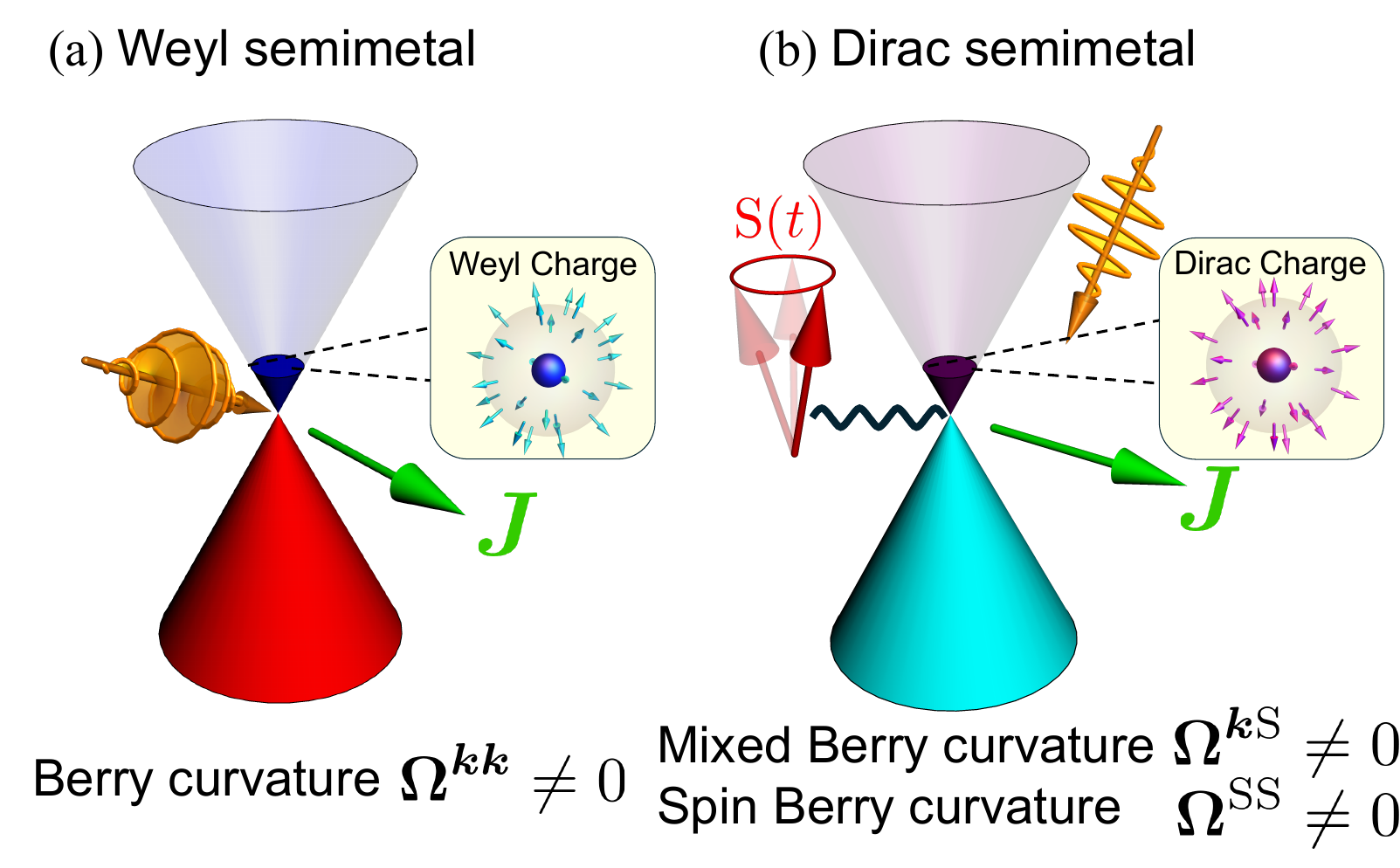}
        \caption{(a) Schematic picture of photocurrent generation driven by circularly polarized light in Weyl semimetals. The green arrow denotes the photocurrent response $J$, and the orange arrow is the circularly polarized light. The photocurrent response reflects the Weyl charge corresponding to the source of the Berry curvature $\bm{\Omega}^{\bm{kk}}$ in the momentum space. (b) Schematic picture of photocurrent generation driven by spin-charge-coupled motive force in Dirac semimetals. The red arrows represent the dynamics of the localized spin moment $\bm{\text{S}}(t)$. The photocurrent response reflects the presence of the Dirac charge corresponding to the source of the mixed Berry curvature $\bm{\Omega^{k\text{S}}}$ or spin Berry curvature $\bm{\Omega^{\text{SS}}}$.}
        \label{topo_charge}
\end{figure}

\textit{Introduction---}
Topological semimetals are characterized by their linear electronic dispersion around the nodes or degenerate points \cite{Burkov2016,Yan_2017,Armitage2018}.
In topological semimetals, the property of the photocurrent response, second-order current response that converts an oscillating light field into a direct current (DC) \cite{Belinicher1980,Baltz1981,Sipe2000}, reflects the topology of the band structure \cite{Nagaosa2020,Liu2020,morimoto2023review,Morimoto_b2016,Chan2017,Yang2018,Parker2019,Avdoshkin2020,Ahn2020, Watanabe2021,Takasan2021,Ikeda2023,Ahn2023,Hsu2023,Ikeda2024,Yoshida2025,Mei2025,Wu2017,Ma2019,Osterhoudt2019}.
Especially Weyl semimetals exhibit the Weyl charge corresponding to the source or sink of the Berry curvature \cite{Murakami2007,Wan2011,Burkov2011,Xu2015}, and the injection current induced by circularly polarized light in the low-frequency regime is shown to be quantized due to the presence of Weyl charges \cite{deJuan2017, Felix2018,deJuan2020,Le2020,Raj2024,Cao2024,Ahn2024} as shown in \figref{topo_charge}(a).
In experiments, the large photocurrent generation induced by the circularly polarized light is observed in Weyl semimetals in the low-frequency regime \cite{Rees2020,Ni2021}. 
On the other hand, such Berry curvature monopole is compensated in Dirac semimetals due to the spin degeneracy.
Thus, the emergent charge carried by the Dirac point remains elusive.

In recent years, photocurrent generation driven by order parameter dynamics has attracted significant attention. 
In the low-frequency regime, collective excitations of order parameters modify the photocurrent response. 
The influence of order parameter dynamics on photocurrent response has been investigated through perturbative analysis \cite{Morimoto2016_exciton, Morimoto2019_electromagnon, Morimoto2021, Toshio2022,Morimoto2024} and real-time simulations \cite{Chan2021, Kaneko2021, Iguchi2024,hattori2025} theoretically. 
Experimentally, photocurrent responses originating from order parameter dynamics have been observed in excitonic insulators \cite{Sotome2019, Sotome2021, Nakamura2024}, ferroelectrics \cite{Okamura2022}, and magnetic insulators \cite{Ogino2024}. 
Photocurrent generation driven by the order parameter dynamics is expected to reflect the band topology in a topological semimetal.

In this work, we discuss the emergence of the Dirac charge and its impact on the photocurrent generation in AFM Dirac semimetals. 
AFM Dirac semimetals have garnered much attention in the context of antiferromagnetic spintronics \cite{Jungwirth_review2016, Baltz2018, Manchon2019}.
In AFM Dirac semimetals, the localized spin system is strongly coupled to the external electric field or current via linear magnetoelectric effects and Edelstein effects \cite{Levitov1985, Edelstein1990, Yanase2014, Zelezny2014} owing to the comparable energy scales of the electronic and localized spin systems.
Therefore, localized spin dynamics is expected to play a crucial role in photocurrent generation in AFM Dirac semimetals.
We introduce the Berry curvature (BC) in the generalized parameter space to characterize the Dirac charge. 
In AFM Dirac semimetals, we consider three types of BC: the BC in momentum space (BCMS), the spin Berry curvature (SBC) in the parameter space of localized spin configurations \cite{Niu1998,Niu1999,Cristopher2017,Simon2022,Lenzing2023}, and the mixed Berry curvature (MBC) \cite{Sundaram1999,Xiao2005,Freimuth2013,Freimuth2014,Hanke2017,Xiao2021,Meguro2025} in the combined parameter space of momentum and spin configurations.
We reveal that the MBC and SBC can be interpreted as a vector field centered on the Dirac point in the momentum space in AFM Dirac semimetals.
We find that the presence of the Dirac charge can be detected via the injection current driven by the spin-charge coupled motive force as shown in \figref{topo_charge}(b).
By evaluating the impact of the Dirac charge on the photocurrent quantitatively through real-time simulation, we clarify that the Dirac charge plays a significant role in the photocurrent generation.
This work opens the hidden property of the Dirac points in the AFM Dirac semimetals.

\textit{Berry curvature in generalized parameter space---}
We introduce the BC in the generalized parameter space to describe the Dirac charges in AFM Dirac semimetals.
We define the Berry connection in the parameter space $\bm{R}$ \cite{Simon1983,Berry1984} as
\begin{align}
    \xi^{X}_{ab}(\bm{R})=\mel{u_a(\bm{R})}{i\partial_{X}}{u_b(\bm{R})},
\end{align}
where $X$ represents the parameter belonging to $\bm{R}$ and the eigenstates $\ket{u_n(\bm{R})}$ of Hamiltonian $\mathcal{H}(\bm{R})$ are characterized by the energy $\epsilon_{n\bm{R}}$ and set of parameter $\bm{R}$.
By using this Berry connection, we can express the quantum geometry of the band $a$ \cite{Resta2011,Ahn2020, Watanabe2021,Ahn2022,Yu2025,Jiang2025,Avdoshkin2025,Mitscherling2025} as $Q_a^{XY}(\bm{R})=\sum_{b\neq a}\xi_{ab}^{X}(\bm{R})\xi_{ba}^{Y}(\bm{R})$,
where the imaginary part $\Omega_a^{XY}(\bm{R})=-2\text{Im}\ Q_a^{XY}(\bm{R})$ represents BC and the real part $g_a^{XY}(\bm{R})=\text{Re}\ Q_a^{XY}(\bm{R})$ denotes the quantum metric.
The BCMS reflects the geometrical property of the electronic band structure in the momentum space. For example, in Weyl semimetals, the BCMS behaves as $\bm{\Omega}^{\bm{kk}}_{a}\propto{\bm{k}}/{\abs{\bm{k}}^3}$ around the Weyl point corresponding to the point charge in the momentum space $\bm{k}$.
Namely, the BCMS originating from the Weyl point is analogous to the classical vector field arising from the point charge.

When we consider the AFM system parameterized by the wavenumber $\bm{k}$ and the localized spin configuration $\bm{\text{S}}$, we can discuss the BCMS, the SBC in $\bm{\text{S}}$-space, and the MBC in the mixed space.
Analogous to BCMS in Weyl semimetals, we can interpret the MBC and SBC as the vector field originating from the Dirac charge in the momentum space in AFM Dirac semimetals.

\textit{Application to AFM Dirac semimetals---}
We consider the BC characterized by the parameters $\bm{R}=(\bm{k,\text{S}})$ in the two-dimensional AFM Dirac semimetal \cite{Smejkal2017,Watanabe2021}, where the Hamiltonian is expressed as 
\begin{align}
    \mathcal{H}=\hat{\mathcal{H}}_{\text{ele}}+\hat{\mathcal{H}}_{\text{exc}}+\mathcal{H}_{\text{mag}}.\label{Kondo}
\end{align}
The first term, $\hat{\mathcal{H}}_{\text{ele}}=\sum_{\bm{k}}\bm{\hat{c}^{\dag}}(\bm{k})\bm{\mathcal{H}}_{\text{ele}}(\bm{k})\bm{\hat{c}}(\bm{k})$ represents the electronic Hamiltonian, where $\bm{\hat{c}}(\bm{k})=(\hat{c}_{A\uparrow}(\bm{k}),\hat{c}_{A\downarrow}(\bm{k}),\hat{c}_{B\uparrow}(\bm{k}),\hat{c}_{B\downarrow}(\bm{k}))^{T}$is a vector of annihilation operators of the electron labeled by sublattice index $\alpha$ ($\alpha = A, B$), spin $\sigma$ ($\sigma=\uparrow, \downarrow$), and wave vector $\bm{k}$. 
The matrix of the Hamiltonian is expressed as
\begin{align}
    \begin{split}
    \bm{\mathcal{H}}_{\text{ele}}(\bm{k})=-2t_1\tau^x\text{cos}\frac{k_x}{2}\text{cos}\frac{k_y}{2}-t_2(\text{cos}k_x+\text{cos}k_y)\\+\lambda\tau_z(\sigma_y\text{sin}k_x-\sigma_x\text{sin}k_y),
    \end{split}
\end{align}
where $t_1$($t_2$) is the nearest-neighbor hopping (next-nearest-neighbor hopping), and $\lambda$ is the staggard spin-orbit coupling. 
Here, the Pauli matrices $\sigma$($\tau$) represent the spin (sublattice) degree of freedom. 
The second term, $\hat{\mathcal{H}}_{\text{exc}}=\sum_{\bm{k}}\bm{\hat{c}^{\dag}}(\bm{k})\bm{\mathcal{H}}_{\text{exc}}(\bm{k,\text{S}})\bm{\hat{c}}(\bm{k})$, represents the interaction between the electronic system and localized spin system, where the matrix is expressed as 
\begin{align}
    \bm{\mathcal{H}}_{\text{exc}}(\bm{k,\text{S}})=J\left[\frac{1+\tau_z}{2}\bm{\sigma}\cdot\bm{\text{S}}_{A}+\frac{1-\tau_z}{2}\bm{\sigma}\cdot\bm{\text{S}}_{B}\right].
\end{align}
Here, $J$ represents the exchange coupling between the electronic spin moment $\bm{\sigma}_{\alpha}$ and the localized spin moment $\bm{\text{S}}_{\alpha}$ on the sublattice site $\alpha$. 
The third term, $\mathcal{H}_{\text{mag}}=\sum_{\alpha=\text{A,B}}K_x(\text{S}_{\alpha}^x)^2$, corresponds to the spin Hamiltonian, where $K_x$ is the coefficient of the easy-axis anisotropy along the $x$-axis.
We introduce this term into the Hamiltonian to stabilize a collinear AFM order, analogous to that observed in CuMnAs \cite{Wadley2016,Tang2016,Smejkal2017,Godinho2018,Linn2023}.

Here, we consider the collinear AFM order $\bm{\text{S}}_{A}=-\bm{\text{S}}_{B}=(1,0,0)$ when we set the parameters as $t_1=1.0,t_2=0.08,\lambda=0.8,J=0.6$.
For this parameter, the Dirac points appear at $k$-points $\bm{\text{D}}_1=[\pi,\text{arccos}(J/\lambda)],\bm{\text{D}}_2=[\pi,\pi-\text{arccos}(J/\lambda)]$ in the electronic band structure.
We define the energy of the lower (upper) Dirac point by $\mu_1$ ($\mu_2$).
These Dirac points are protected by the nonsymmorphic mirror symmetry \cite{Tang2016,Smejkal2017,Yang2017}.

We define the symmetry-adapted basis for the localized spin system as $\mathrm{M}^{a}(t) = \frac{1}{2}\qty(\mathrm{S}_{A}^{a}(t) + \mathrm{S}_{B}^{a}(t)), \mathrm{L}^{a}(t) = \frac{1}{2}\qty(\mathrm{S}_{A}^{a}(t) - \mathrm{S}_{B}^{a}(t))$, and $\mathrm{M}^{z}$ and $\mathrm{L}^{y}$ can be excited by the external electric field along the $x$-diretion.
We can access to the electron's spin operators $\mathrm{M}_{\sigma}^{z}=\sigma_{z}\tau_{0}$ and $\mathrm{L}_{\sigma}^{y}=\sigma_{y}\tau_{z}$ conjugate with the light-driven localized spin dynamics $\mathrm{M}^{z}$ and $\mathrm{L}^{y}$, where $\tau_{0}$ represents the identity in the sublattice degree of freedom.
Therefore, we discuss the BC $\Omega^{XY}$ for the effective Dirac Hamiltonian in the case of $X,Y=k^{x},\mathrm{M}^{z}$ or $\mathrm{L}^{y}$.

Following Ref.\cite{Watanabe2021}, we consider the effective Dirac Hamiltonian around the Dirac points $\bm{\text{D}}_1$ and $\bm{\text{D}}_2$ given by 
\begin{align}
    H_{\text{eff}}(\bm{k};s_z)=v_0k_y+a_1k_y\sigma_x-a_2k_x\sigma_y+wk_x\tau_x+\Delta,
\end{align}
where the parameters are defined as $v_0=t_2\mathrm{cos}k_0,a_1=\lambda\mathrm{sin}k_0s_z,a_2=\lambda, w=t_1\mathrm{cos}\left(\frac{\pi/2+s_zk_0}{2}\right),\Delta=t_2\mathrm{sin}k_0s_z$.
Here, $s_z=\pm1$ corresponds to the Dirac nodes at $\bm{\text{D}}_1$ and $\bm{\text{D}}_2$, respectively.
The eigenenergy of the Dirac Hamiltonian can be expressed as $\epsilon_{\bm{k}\pm;s_z}=\abs{\bm{\rho}}\left(\frac{v_0}{|a_1|}\mathrm{sin}\theta\pm1\right)+\Delta$, where $\bm{\rho}=(\rho_x,\rho_y)$ satisfies $\abs{\bm{\rho}}=\sqrt{a_1^2k_y^2+(a_2^2+w^2)k_x^2},\  \rho_x=\sqrt{a_2^2+w^2}k_x,\ \rho_y=|a_1|k_y$.

We derive the analytical expression for the BC of this effective Dirac Hamiltonian $\Omega^{XY}\equiv \Omega_{-}^{XY}=-\Omega_{+}^{XY}$ as 
\begin{align}
    \Omega^{\text{M}^zk^x}&={\text{sgn}(a_1)a_2}\frac{\rho_y}{\rho^3},\ \Omega^{\text{L}^y\text{M}^z}={\text{sgn}(a_1)}\frac{\rho_y}{\rho^3},
\end{align}
where the index $-$ $(+)$ denotes the lower band (upper band) of the Dirac dispersion.
Other components of the BC vanish owing to the $\mathcal{PT}$-symmetry.
The detailed derivation is summarized in Supplemental Materials (SM) \cite{supple}.
From these expressions for the BC, we can interpret the BC as the $\rho_y$-element of the vector field originating from the Dirac point scaling as ${\bm{\rho}}/{\abs{\bm{\rho}}^3}$.
These behaviors of the BC correspond to the vector field from the point charge in three dimensions.
In this sense, we can regard the Dirac point as the Dirac charge corresponding to the sink or source of the MBC and SBC in the momentum space.

\textit{Detection of Dirac charge---}
To detect the Dirac charge of AFM Dirac semimetals, we utilize the injection current originating from the spin-charge-coupled motive force reflecting the BC. 
We can express the generalized injection current as
\begin{align}\label{injection_current}
    J_{\text{Inj}}^{\mu;XY}(0,\omega_{\text{p}})&=\chi_{\text{Inj}}^{\mu;XY}(0;-\omega_{\text{p}},\omega_{\text{p}})X(-\omega_{\text{p}})Y(\omega_{\text{p}})\nonumber\\
    &+\chi_{\text{Inj}}^{\mu;XY}(0;\omega_{\text{p}},-\omega_{\text{p}})X(\omega_{\text{p}})Y(-\omega_{\text{p}}),
\end{align}
where $ X(\omega_{\text{p}})$ and $Y(\omega_{\text{p}})$ correspond to the external fields such as the electric field or the light field driven localized spin dynamics with frequency $\omega_{\text{p}}$.
The general formula for the injection current conductivity $\chi_{\text{Inj}}^{\mu;XY}(0;-\omega_{\text{p}},\omega_{\text{p}})$ \cite{Iguchi2024,hattori2025} is expressed as
\begin{align}\label{formula_injection}
    &\chi_{\text{Inj}}^{\mu;XY}(0;-\omega_{\text{p}},\omega_{\text{p}})\nonumber\\ 
        &= \pi\tau\int \dfrac{d\vb*{k}}{(2\pi)^{d}}\sum_{a\neq b}\Delta_{ab}^{\mu}{X}_{ab}Y_{ba}f_{ab}\delta(\omega_{\text{p}}-\epsilon_{ba}).
\end{align}
Here, $\Delta_{ab}^{\mu} = \partial_{\mu} (\epsilon_{a\bm{k}} - \epsilon_{b\bm{k}})$ is the velocity difference matrix, $X_{ab}, Y_{ba}$ represent the matrix element of either the Berry connection $\vb*{\xi}_{ab} = \mel{u_{a\vb*{k}}}{i\nabla_{\vb*{k}}}{u_{b\vb*{k}}}$ in the momentum space or the spin operator $\bm{\sigma}_{ab}=\mel{u_a(\bm{R})}{\bm{\sigma}}{u_b(\bm{R})}$ conjugate with the external field $ X(\omega_{\text{p}})$ and $Y(\omega_{\text{p}})$, the quantity $f_{ab} = f_a - f_b$ denotes the difference in the Fermi distribution function, where $f_a = f(\epsilon_{a\bm{k}})$, and $\epsilon_{ba} = \epsilon_{b\bm{k}} - \epsilon_{a\bm{k}}$ represents the energy difference between bands $a$ and $b$. 
We can derive the generalized injection current conductivity for the effective Dirac Hamiltonian at zero temperature as
\begin{align}
    &\chi_{\text{Inj};s_z}^{\mu;XY}(0;-\omega_{\text{p}},\omega_{\text{p}})\nonumber\\
    &=-\frac{\tau c_{\mu}(i\omega_{\text{p}})^{\delta_X}(-
    i\omega_{\text{p}})^{\delta_Y}}{4\pi\abs{a_1}\sqrt{a_2^2+w^2}}\int_{\substack{\abs*{\rho}=\frac{\omega_{\text{p}}}{2}\\ f_{ab}=1}} d\bm{\rho}\frac{\rho_{\mu}}{\abs{\bm{\rho}}}Q_{s_z}^{R^XR^Y},
\end{align}
where the integral represents the integration under the condition $\abs{\bm{\rho}}=\omega_{\text{p}}/2$ and $f_{ab}=1$, $Q_{s_z}^{R^XR^Y}$ denotes the quantum geometry in the Dirac node $s_z$, and $c_{\mu}$ is defined by $\rho_{\mu}=c_{\mu}k_{\mu}$.
Here, $R^X$ represents the parameter related to the physical quantity $X$, and $\delta_X$ is defined such that $\delta_X=0$ when $R^X\in\bm{k}$, and $\delta_{X}=1$ when $R^X\in\bm{\text{S}}$.
Therefore, the injection current reflects the source or sink of quantum geometry $Q_{s_z}^{R^XR^Y}$.

We define the Dirac charge $C^{XY}_{\text{D}}$ as the source or sink of the BC as
\begin{align}
    C^{XY}_{\text{D}}=\frac{1}{2\pi}\omega_{\text{p}}^{\delta_X+\delta_Y}\oint_{\rho=\frac{\omega_{\text{p}}}{2}}d\bm{S}\cdot\bm{\Omega}^{R^XR^Y},\label{Dirac_charge_eq}
\end{align}
where the integral represents the integration under $\abs{\bm{\rho}}=\omega_{\text{p}}/2$ and $d\bm{S}$ represents the oriented surface element normal to surface and $\bm{\Omega}^{R^XR^Y}=b^{R^XR^Y}\bm{\rho}/\abs{\bm{\rho}}^3$ is the vector field characterized by the coefficient $b^{R^XR^Y}$ in $\Omega^{R^XR^Y}$.
In two-dimensional systems, $C^{XY}_{\text{D}}$ characterized by $\delta_X+\delta_Y=1$ is a constant value independent of the integration radius.
Therefore, the Dirac charge of the MBC corresponds to the point charge in a two-dimensional system.
In contrast, the Dirac charge of the SBC is the source or sink depending on the integration radius.
We can also extend the concept of the Dirac charge to three-dimensional Dirac semimetals \cite{Tang2016} easily.

\begin{figure}
        \centering
        \includegraphics[width=\linewidth]{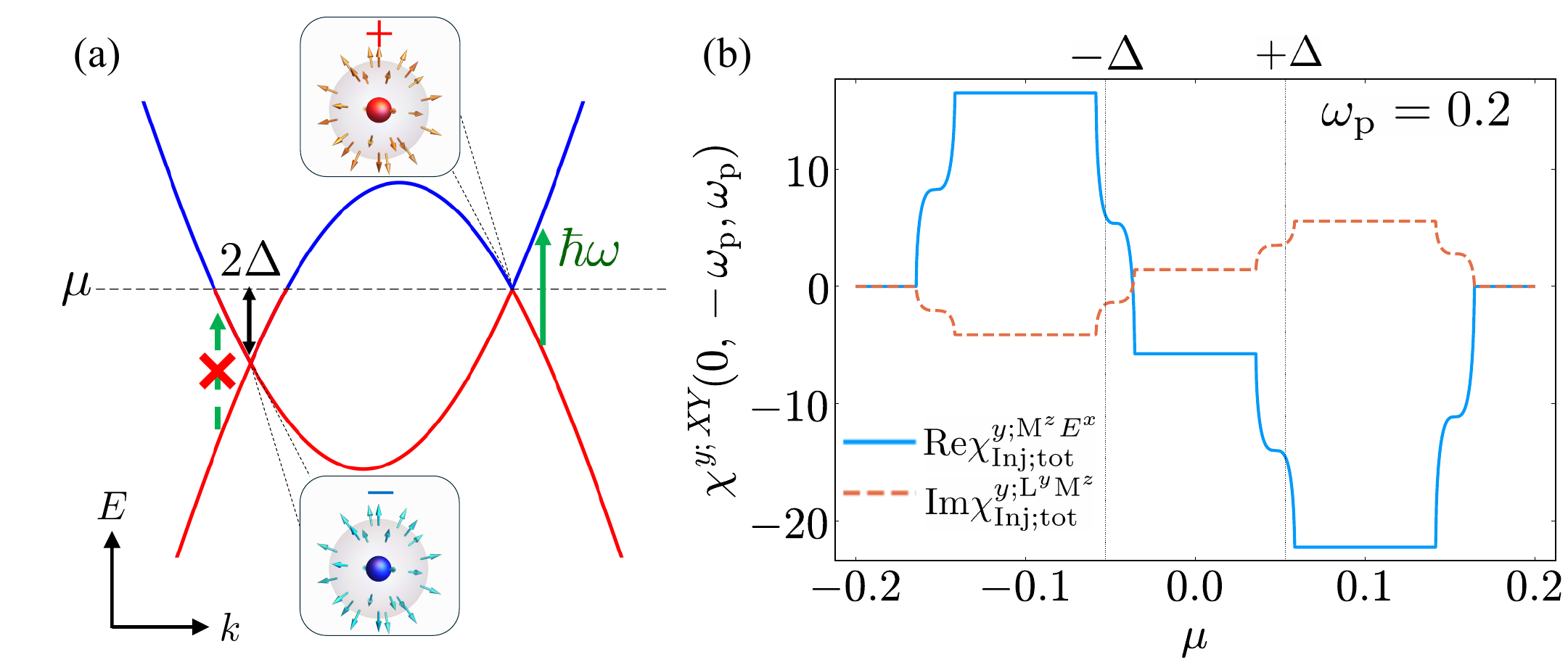}
        \caption{(a) Schematic illustration of photocurrent generation in Dirac semimetals when two Dirac nodes are at different energies. The two Dirac nodes have different signs of the Dirac charge $\pm C_{\text{D}}$. The energy difference $2\Delta$ between the two Dirac nodes prevents the cancellation of the contributions to the photocurrent response driven by the external field with frequency $\hbar\omega$ from these Dirac charges owing to the Pauli blocking. (b) Chemical potential dependence of injection current conductivity $\chi_{\text{Inj;tot}}^{\mu;XY}(0;-\omega_{\text{p}},\omega_{\text{p}})$ obtained from the analytical expression in the effective Dirac model when $\omega_{\text{p}}=0.2$ and $\tau=100.0$. The blue solid line represents $\text{Re}\ \chi_{\text{Inj;tot}}^{y;\text{M}^zE^{x}}$ and the orange dashed line denotes $\text{Im}\ \chi_{\text{Inj;tot}}^{y;\text{L}^y\text{M}^z}$. The vertical dot lines denote the energy of the Dirac points $\pm\Delta$. }
        \label{analytical_injection_fig}
\end{figure}

By performing the symmetry analysis of the photocurrent in the AFM Dirac semimetal \cite{supple}, the nonlinear transverse photocurrent conductivity driven by the external electric field along the $x$-axis and the localized spin dynamics can be finite.
Therefore, we discuss the photocurrent generation along the $y$-direction $\chi^{y;XY}$ in the case of $X,Y=E^{x},\mathrm{M}^{z}$ or $\mathrm{L}^{y}$.
We derive the analytical expression for the injection current conductivity $\chi_{\mathrm{Inj};s_z}^{y;XY}(0;-\omega_{\text{p}},\omega_{\text{p}})$ of the Dirac node $s_z$ as
\begin{align}
    \chi_{\mathrm{Inj};s_z}^{y;\mathrm{M}^zE^x}(0;-\omega_{\text{p}},\omega_{\text{p}}) &=-\frac{\tau a_2\text{sgn}(a_1)}{2\pi\sqrt{a_2^2+w^2}}\left[\frac{\theta}{2}-\frac{1}{4}\mathrm{sin}2\theta\right]^{\theta_+}_{\theta_-},\label{inj_MzE}\\
    \chi_{\mathrm{Inj};s_z}^{y;\mathrm{L}^y\mathrm{M}^z}(0;-\omega_{\text{p}},\omega_{\text{p}}) &=\frac{i\tau\omega_{\text{p}}\text{sgn}(a_1)}{2\pi\sqrt{a_2^2+w^2}}\left[\frac{\theta}{2}-\frac{1}{4}\mathrm{sin}2\theta\right]^{\theta_+}_{\theta_-},\label{inj_LyMz}
\end{align}
where we define $\theta_{\pm}$ as
\begin{align}
    \theta_{\pm}=\begin{cases}
        \frac{\pi}{2} & A_{\pm}\ge 1\\
        \text{arcsin}(A_{\pm})& -1<A_{\pm}<1\\
        -\frac{\pi}{2}& A_{\pm}\le -1
    \end{cases}.
\end{align}
Here, we introduce $A_{\pm}=\frac{|a_1|}{v_0}\left(\pm1-\frac{2(-\mu+\Delta)}{\omega_{\text{p}}}\right)$.
In the case of $\theta_{\pm}=\pm\frac{\pi}{2}$, we can express the injection current conductivity coming from the BC using the Dirac charge $C_{\text{D};s_z}^{XY}$ in the Dirac node $s_z$ as
\begin{align}
    \chi_{\text{IBC};s_z}^{\mu;XY}(0;-\omega_{\text{p}},\omega_{\text{p}})=\frac{i^{\delta_X+1}
    (-i)^{\delta_Y}\tau C_{\text{D};s_z}^{XY}}{8\sqrt{a_2^2+w^2}}.
\end{align}
Owing to this fact, we can detect the Dirac charge via the injection current driven by the external light and the localized spin dynamics.
The sign of the injection current conductivity differs between the two Dirac nodes owing to the sign of $a_1$. 
Therefore, the contribution to the injection current from the different nodes cancels with each other as shown in \figref{analytical_injection_fig}(a).
However, the energy difference $2\Delta$ between the two Dirac nodes prevents the cancellation of the contributions to the photocurrent response from these Dirac charges owing to the Pauli blocking.

By using the formulas in \Eqref{inj_MzE} and \eqref{inj_LyMz}, we plot the chemical potential dependence of the total injection current conductivity expressed as $\chi_{\text{Inj;tot}}^{y;XY}(0;-\omega_{\text{p}},\omega_{\text{p}})=\chi_{\text{Inj};+1}^{y;XY}(0;-\omega_{\text{p}},\omega_{\text{p}})+\chi_{\text{Inj};-1}^{y;XY}(0;-\omega_{\text{p}},\omega_{\text{p}})$ in \figref{analytical_injection_fig}(b).
In \figref{analytical_injection_fig}(b), both injection current conductivities $\text{Re}\ \chi_{\text{Inj;tot}}^{y;\text{M}^zE^x}(0;-\omega_{\text{p}},\omega_{\text{p}})$ and $\text{Im}\ \chi_{\text{Inj;tot}}^{y;\text{M}^z\text{L}^y}(0;-\omega_{\text{p}},\omega_{\text{p}})$ exhibit the flat-topped shapes reflecting the presence of the Dirac point around $\pm\Delta$ corresponding to the energy level of the Dirac points.
Especially, the value of the injection current conductivity at the peak reflects the Dirac charge defined in \Eqref{Dirac_charge_eq}.
We note that other injection current conductivities coming from the quantum geometry do not show this behavior \cite{supple}.

\textit{Real-time simulation---}
To evaluate the photocurrent generation from the spin-charge-coupled dynamics quantitatively, we solve the two-coupled equations of motion, the von Neumann equation and the Landau-Lifshitz-Gilbert (LLG) equation following Ref.\cite{Ono2021, ono2023, Iguchi2024, hattori2024, hattori2025}. 
For the electronic system, we solve the von Neumann equation \cite{Yue2022,Murakami2022,Bharti2023,Bharti2024} for the single-particle density matrix (SPDM) $\rho_{\alpha \beta}^{\sigma \sigma^{\prime}}(\bm{k}) = \expecval{\hat{c}_{\beta\sigma^{\prime}}^{\dagger}(\bm{k})\hat{c}_{\alpha\sigma}(\bm{k})}$ expressed as
\begin{align}
    \begin{split}
            \pdv{\vb*{\rho}(\bm{k}, t)}{t} = -i\qty[\vb*{H}(\bm{k},t),\, \vb*{\rho}(\bm{k}, t)] - \bm{E}(t)\cdot\frac{\partial\vb*{\rho}(\bm{k}, t)}{\partial \bm{k}} \\
        - \frac{1}{\tau}(\vb*{\rho}(\bm{k}, t) - \vb*{\rho}_{\text{eq}}(\bm{k})).
        \label{vonNeumann}
        \end{split}
\end{align}
Here, $\tau$ represents the relaxation time of the electronic system, $\bm{E}(t)$ denotes the external electric field and $\vb*{H}(\bm{k},t)$ is the time-dependent electronic Hamiltonian at each $\bm{k}$ point defined as $\hat{\mathcal{H}}_{\mathrm{ele}} + \hat{\mathcal{H}}_{\mathrm{exc}} = \sum_{\bm{k}}\bm{\hat{c}^{\dag}}(\bm{k})\vb*{H}(\bm{k},t)\bm{\hat{c}}(\bm{k})$. 

On the other hand, we solve the LLG equation for the localized spin moment $\bm{\text{S}}_{\alpha}$ described as 
\begin{align}
    \dfrac{d\boldsymbol{\mathrm{S}}_{\alpha}}{dt} &= \dfrac{1}{1 + \alpha_{G}^{2}}\left( \boldsymbol{\mathrm{h}}_{\alpha}^{\mathrm{eff}} \times \boldsymbol{\mathrm{S}}_{\alpha} + \alpha_{G} \boldsymbol{\mathrm{S}}_{\alpha} \times \left( \boldsymbol{\mathrm{S}}_{\alpha} \times\boldsymbol{\mathrm{h}}_{\alpha}^{\mathrm{eff}} \right)\right) \label{LLG},
\end{align}
where $\boldsymbol{\mathrm{h}}_{\alpha}^{\mathrm{eff}} = -J\langle \boldsymbol{\sigma}_{\alpha} \rangle + \dfrac{\delta\mathcal{H}_{\text{spin}}}{\delta \boldsymbol{\mathrm{S}}_{\alpha}}$ is the effective magnetic field on the localized spin moment on the sublattice $\alpha$ and $\alpha_G$ is the Gilbert damping coefficient.
In the following calculation, we set the $k$-mesh of the Brillouin zone to $1000\times1000$, and use $K_x=0.1$, $\tau=100$, $\alpha_G=0.01$, and inverse temperature $\beta=500.0$.
Notably, the injection current is linearly proportional to the relaxation time, and $\tau = 100$ is sufficiently large for the injection currents to dominate the photocurrent conductivity in the finite-frequency regime. 
Under these conditions, we discuss the $\mu$-dependence of the injection current through real-time simulation of the Kondo lattice Hamiltonian in \Eqref{Kondo}.

Using this scheme of real-time simulation, we evaluate the transverse photocurrent response described as
\begin{align}
    J^{y}(\omega=0;\omega_{\text{p}})&=\sigma^{yxx}(0;-\omega_{\text{p}},\omega_{\text{p}})E^{x}(-\omega_{\text{p}})E^{x}(\omega_{\text{p}})\nonumber\\
    &+\sigma^{yxx}(0;\omega_{\text{p}},-\omega_{\text{p}})E^{x}(\omega_{\text{p}})E^{x}(-\omega_{\text{p}}),
\end{align}
where $J^{y}(\omega=0;\omega_{\text{p}})$ is the DC component of the current response under the external electric field with frequency $\omega_{\text{p}}$.
We define the photocurrent conductivity as $\sigma^{yxx}(0;\omega_{\text{p}})=(\sigma^{yxx}(0;-\omega_{\text{p}},\omega_{\text{p}})+\sigma^{yxx}(0;\omega_{\text{p}},-\omega_{\text{p}}))/2$ in this study.
To understand the behavior of the photocurrent response, we decompose the photocurrent response $\sigma^{yxx}_{\text{w}}(0;\omega_p)$, including the effects of both the electric field and localized spin dynamics, into three components as
\begin{align}
    \sigma^{yxx}_{\text{w}}(0;\omega_p)=\sigma^{yxx}_{\text{wo}}(0;\omega_p)+\sigma^{yxx}_{\text{col-}E}(0;\omega_p)+\sigma^{yxx}_{\text{col-col}}(0;\omega_p),
\end{align}
where $\sigma^{yxx}_{\text{wo}}(0;\omega_p)$ is the photocurrent coming from the electric field effect, $\sigma^{yxx}_{\text{col-}E}(0;\omega_p)$ is the photocurrent generation from the interference between the localized spin dynamics and electric field, and $\sigma^{yxx}_{\text{col-col}}(0;\omega_p)$ arises from the localized spin dynamics.

\begin{figure}[t]
        \centering
        \includegraphics[width=\linewidth]{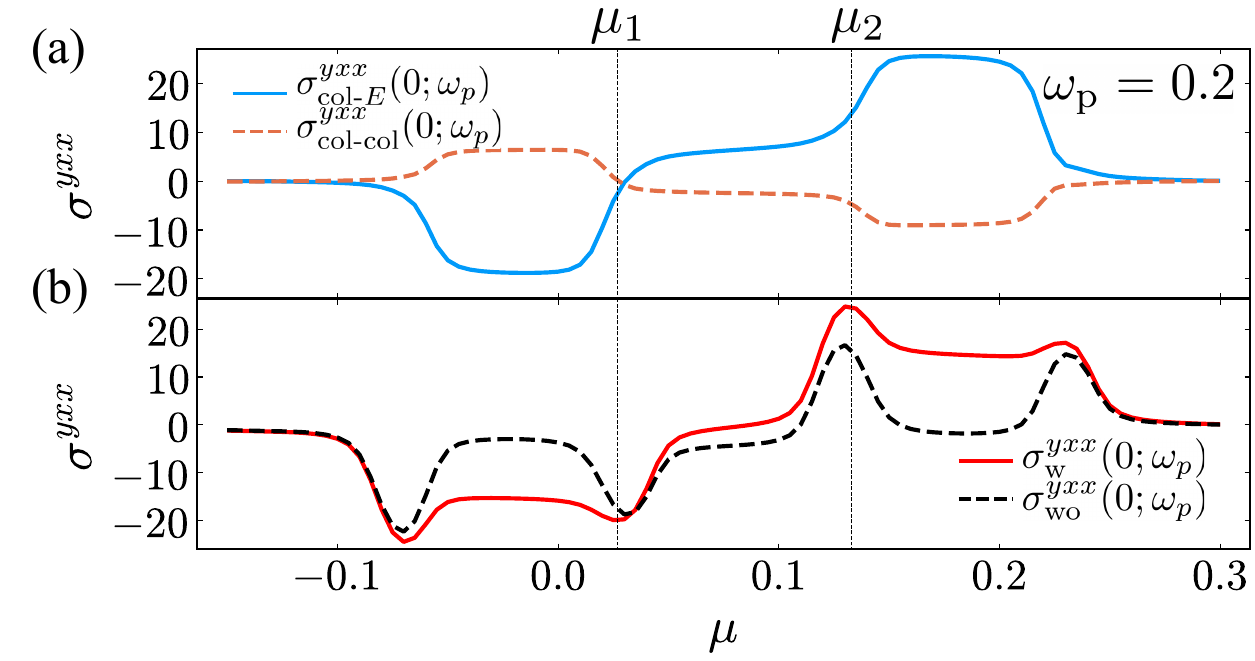}
        \caption{$\mu$-dependence of photocurrent conductivity at resonance frequency $\omega_{\text{p}}=0.2$ obtained by real-time simulation when $\tau=100$.
        (a) $\mu$-dependence of $\sigma_{\text{col-}E}(0,\omega_{\text{p}}=0.2)$ and $\sigma_{\text{col-col}}(0,\omega_{\text{p}}=0.2)$. The blue solid line represents $\sigma_{\text{col-}E}(0,\omega_{\text{p}}=0.2)$ and the orange dashed line denotes $\sigma_{\text{col-col}}(0,\omega_{\text{p}}=0.2)$. (b) $\mu$-dependence of $\sigma_{\text{w}}(0,\omega_{\text{p}}=0.2)$ and $\sigma_{\text{wo}}(0,\omega_{\text{p}}=0.2)$. The red solid line represents $\sigma_{\text{w}}(0,\omega_{\text{p}}=0.2)$ and the black dashed line denotes $\sigma_{\text{wo}}(0,\omega_{\text{p}}=0.2)$. The chemical potentials $\mu_1$ and $\mu_2$ are represented by the vertical dotted lines.}
        \label{chemical_fig}
\end{figure}

We present the $\mu$-dependence of the photocurrent conductivity $\sigma^{yxx}_{\text{col-}E}(0;\omega_{\text{p}})$ and $\sigma^{yxx}_{\text{col-col}}(0;\omega_{\text{p}})$, at the resonance frequency of the localized spin system $\omega_{\text{p}} = 0.2$ in \figref{chemical_fig}(a).
The $\mu$-dependence of $\sigma^{yxx}_{\text{col-}E}(0;\omega_{\text{p}})$ and $\sigma^{yxx}_{\text{col-col}}(0;\omega_{\text{p}})$ in \figref{chemical_fig}(a) is consistent with the result obtained from the analytical calculation in \figref{analytical_injection_fig}(b), and they exhibit the peak structure reflecting the Dirac charges around the Dirac points.
We show the $\mu$-dependence of $\sigma^{yxx}_{\text{w}}(0;\omega_p)$ and $\sigma^{yxx}_{\text{wo}}(0;\omega_p)$ at $\omega_{\text{p}}$ = 0.2 in \figref{chemical_fig}(b).
We observe the narrow peaks around $\mu=-0.05$ and $\mu=0.2$ in the $\mu$-dependence of $\sigma^{yxx}_{\text{wo}}(0;\omega_{\text{p}})$, which reflect the $\mu$-dependence of the injection current conductivity induced by the linearly polarized light.
This contribution comes from the quantum metric in momentum space \cite{Zhang2019}.
In contrast, $\sigma^{yxx}_{\text{w}}(0;\omega_{\text{p}})$, which incorporates the influence of the Dirac charges, exhibits the broad peaks owing to the photocurrent contributions from localized spin dynamics, $\sigma^{yxx}_{\text{col-}E}(0;\omega_{\text{p}})$ and $\sigma^{yxx}_{\text{col-col}}(0;\omega_{\text{p}})$. 
These contributions from the Dirac charge are comparable to $\sigma^{yxx}_{\text{wo}}(0;\omega_{\text{p}})$ and play a significant role in the photocurrent generation in the broadband filling.

\textit{Summary---}
In this work, we discussed the emergence of the Dirac charge and its detection via the photocurrent generation in the AFM Dirac semimetals.
First, we found that the BC in the generalized parameter space can be regarded as a vector field centered on the Dirac points in the momentum space.
Second, we revealed that the injection current is characterized by the source or sink of the MBC and SBC around the Dirac points, and the Dirac charge can be detected experimentally via the injection current driven by the spin-charge-coupled motive force. 
Finally, we clarified that the contribution from the Dirac charge to the photocurrent conductivity is comparable to that coming from the electric field effect and plays a significant role in the photocurrent generation by performing the real-time simulation of the Kondo lattice model.
The Dirac charge can be detected via the injection current in electrically switchable AFM Dirac semimetals such as CuMnAs \cite{Wadley2016,Tang2016,Smejkal2017,Godinho2018,Linn2023}.
We note that it is straightforward to extend the concept of the Dirac charge to other Fermion-Boson coupled systems, such as electron-phonon coupling systems \cite{Okamura2022,Morimoto2024}.

\textit{Acknowledgement---}
K.H. thanks Manabu Sato and Ming-Chun Jiang for helpful discussion.
This work is supported by Grant-in-Aid for Scientific Research from JSPS, KAKENHI Grant No.~JP23K13058 (H.W.), No.~JP24K00581 (H.W.), No.~JP25H02115 (H.W.), No.~21H04990 (R.A.), JST-CREST No.~JPMJCR23O4(R.A.), No.~25H01246 (R.A.), No. 25H01252 (R.A.), JST-ASPIRE No.~JPMJAP2317 (R.A.), JST-Mirai No.~JPMJMI20A1 (R.A.), JST SPRING, Grant Number JPMJSP2108 (K.H.). K.H. was supported by the Program for Leading Graduate Schools (MERIT-WINGS).
This work was supported by the RIKEN TRIP initiative (RIKEN Quantum, Advanced General Intelligence for Science Program, Many-body Electron Systems).

\bibliography{refs}

\begin{thebibliography}{95}%
\makeatletter
\providecommand \@ifxundefined [1]{%
 \@ifx{#1\undefined}
}%
\providecommand \@ifnum [1]{%
 \ifnum #1\expandafter \@firstoftwo
 \else \expandafter \@secondoftwo
 \fi
}%
\providecommand \@ifx [1]{%
 \ifx #1\expandafter \@firstoftwo
 \else \expandafter \@secondoftwo
 \fi
}%
\providecommand \natexlab [1]{#1}%
\providecommand \enquote  [1]{``#1''}%
\providecommand \bibnamefont  [1]{#1}%
\providecommand \bibfnamefont [1]{#1}%
\providecommand \citenamefont [1]{#1}%
\providecommand \href@noop [0]{\@secondoftwo}%
\providecommand \href [0]{\begingroup \@sanitize@url \@href}%
\providecommand \@href[1]{\@@startlink{#1}\@@href}%
\providecommand \@@href[1]{\endgroup#1\@@endlink}%
\providecommand \@sanitize@url [0]{\catcode `\\12\catcode `\$12\catcode `\&12\catcode `\#12\catcode `\^12\catcode `\_12\catcode `\%12\relax}%
\providecommand \@@startlink[1]{}%
\providecommand \@@endlink[0]{}%
\providecommand \url  [0]{\begingroup\@sanitize@url \@url }%
\providecommand \@url [1]{\endgroup\@href {#1}{\urlprefix }}%
\providecommand \urlprefix  [0]{URL }%
\providecommand \Eprint [0]{\href }%
\providecommand \doibase [0]{https://doi.org/}%
\providecommand \selectlanguage [0]{\@gobble}%
\providecommand \bibinfo  [0]{\@secondoftwo}%
\providecommand \bibfield  [0]{\@secondoftwo}%
\providecommand \translation [1]{[#1]}%
\providecommand \BibitemOpen [0]{}%
\providecommand \bibitemStop [0]{}%
\providecommand \bibitemNoStop [0]{.\EOS\space}%
\providecommand \EOS [0]{\spacefactor3000\relax}%
\providecommand \BibitemShut  [1]{\csname bibitem#1\endcsname}%
\let\auto@bib@innerbib\@empty
\bibitem [{\citenamefont {Burkov}(2016)}]{Burkov2016}%
  \BibitemOpen
  \bibfield  {author} {\bibinfo {author} {\bibfnamefont {A.~A.}\ \bibnamefont {Burkov}},\ }\bibfield  {title} {\bibinfo {title} {Topological semimetals},\ }\href {https://doi.org/10.1038/nmat4788} {\bibfield  {journal} {\bibinfo  {journal} {Nature Materials}\ }\textbf {\bibinfo {volume} {15}},\ \bibinfo {pages} {1145} (\bibinfo {year} {2016})}\BibitemShut {NoStop}%
\bibitem [{\citenamefont {Yan}\ and\ \citenamefont {Felser}(2017)}]{Yan_2017}%
  \BibitemOpen
  \bibfield  {author} {\bibinfo {author} {\bibfnamefont {B.}~\bibnamefont {Yan}}\ and\ \bibinfo {author} {\bibfnamefont {C.}~\bibnamefont {Felser}},\ }\bibfield  {title} {\bibinfo {title} {Topological materials: Weyl semimetals},\ }\href {https://doi.org/10.1146/annurev-conmatphys-031016-025458} {\bibfield  {journal} {\bibinfo  {journal} {Annual Review of Condensed Matter Physics}\ }\textbf {\bibinfo {volume} {8}},\ \bibinfo {pages} {337} (\bibinfo {year} {2017})}\BibitemShut {NoStop}%
\bibitem [{\citenamefont {Armitage}\ \emph {et~al.}(2018)\citenamefont {Armitage}, \citenamefont {Mele},\ and\ \citenamefont {Vishwanath}}]{Armitage2018}%
  \BibitemOpen
  \bibfield  {author} {\bibinfo {author} {\bibfnamefont {N.~P.}\ \bibnamefont {Armitage}}, \bibinfo {author} {\bibfnamefont {E.~J.}\ \bibnamefont {Mele}},\ and\ \bibinfo {author} {\bibfnamefont {A.}~\bibnamefont {Vishwanath}},\ }\bibfield  {title} {\bibinfo {title} {Weyl and dirac semimetals in three-dimensional solids},\ }\href {https://doi.org/10.1103/RevModPhys.90.015001} {\bibfield  {journal} {\bibinfo  {journal} {Review of Modern Physics}\ }\textbf {\bibinfo {volume} {90}},\ \bibinfo {pages} {015001} (\bibinfo {year} {2018})}\BibitemShut {NoStop}%
\bibitem [{\citenamefont {Belinicher}\ and\ \citenamefont {Sturman}(1980)}]{Belinicher1980}%
  \BibitemOpen
  \bibfield  {author} {\bibinfo {author} {\bibfnamefont {V.~I.}\ \bibnamefont {Belinicher}}\ and\ \bibinfo {author} {\bibfnamefont {B.~I.}\ \bibnamefont {Sturman}},\ }\bibfield  {title} {\bibinfo {title} {The photogalvanic effect in media lacking a center of symmetry},\ }\href {https://doi.org/10.1070/PU1980v023n03ABEH004703} {\bibfield  {journal} {\bibinfo  {journal} {Soviet Physics Uspekhi}\ }\textbf {\bibinfo {volume} {23}},\ \bibinfo {pages} {199} (\bibinfo {year} {1980})}\BibitemShut {NoStop}%
\bibitem [{\citenamefont {von Baltz}\ and\ \citenamefont {Kraut}(1981)}]{Baltz1981}%
  \BibitemOpen
  \bibfield  {author} {\bibinfo {author} {\bibfnamefont {R.}~\bibnamefont {von Baltz}}\ and\ \bibinfo {author} {\bibfnamefont {W.}~\bibnamefont {Kraut}},\ }\bibfield  {title} {\bibinfo {title} {Theory of the bulk photovoltaic effect in pure crystals},\ }\href {https://doi.org/10.1103/PhysRevB.23.5590} {\bibfield  {journal} {\bibinfo  {journal} {Physical Review B}\ }\textbf {\bibinfo {volume} {23}},\ \bibinfo {pages} {5590} (\bibinfo {year} {1981})}\BibitemShut {NoStop}%
\bibitem [{\citenamefont {Sipe}\ and\ \citenamefont {Shkrebtii}(2000)}]{Sipe2000}%
  \BibitemOpen
  \bibfield  {author} {\bibinfo {author} {\bibfnamefont {J.~E.}\ \bibnamefont {Sipe}}\ and\ \bibinfo {author} {\bibfnamefont {A.~I.}\ \bibnamefont {Shkrebtii}},\ }\bibfield  {title} {\bibinfo {title} {Second-order optical response in semiconductors},\ }\href {https://doi.org/10.1103/PhysRevB.61.5337} {\bibfield  {journal} {\bibinfo  {journal} {Physical Review B}\ }\textbf {\bibinfo {volume} {61}},\ \bibinfo {pages} {5337} (\bibinfo {year} {2000})}\BibitemShut {NoStop}%
\bibitem [{\citenamefont {Nagaosa}\ \emph {et~al.}(2020)\citenamefont {Nagaosa}, \citenamefont {Morimoto},\ and\ \citenamefont {Tokura}}]{Nagaosa2020}%
  \BibitemOpen
  \bibfield  {author} {\bibinfo {author} {\bibfnamefont {N.}~\bibnamefont {Nagaosa}}, \bibinfo {author} {\bibfnamefont {T.}~\bibnamefont {Morimoto}},\ and\ \bibinfo {author} {\bibfnamefont {Y.}~\bibnamefont {Tokura}},\ }\bibfield  {title} {\bibinfo {title} {Transport, magnetic and optical properties of weyl materials},\ }\href {https://doi.org/10.1038/s41578-020-0208-y} {\bibfield  {journal} {\bibinfo  {journal} {Nature Reviews Materials}\ }\textbf {\bibinfo {volume} {5}},\ \bibinfo {pages} {621} (\bibinfo {year} {2020})}\BibitemShut {NoStop}%
\bibitem [{\citenamefont {Liu}\ \emph {et~al.}(2020)\citenamefont {Liu}, \citenamefont {Xia}, \citenamefont {Xiao}, \citenamefont {Garc{\'i}a~de Abajo},\ and\ \citenamefont {Sun}}]{Liu2020}%
  \BibitemOpen
  \bibfield  {author} {\bibinfo {author} {\bibfnamefont {J.}~\bibnamefont {Liu}}, \bibinfo {author} {\bibfnamefont {F.}~\bibnamefont {Xia}}, \bibinfo {author} {\bibfnamefont {D.}~\bibnamefont {Xiao}}, \bibinfo {author} {\bibfnamefont {F.~J.}\ \bibnamefont {Garc{\'i}a~de Abajo}},\ and\ \bibinfo {author} {\bibfnamefont {D.}~\bibnamefont {Sun}},\ }\bibfield  {title} {\bibinfo {title} {Semimetals for high-performance photodetection},\ }\href {https://doi.org/10.1038/s41563-020-0715-7} {\bibfield  {journal} {\bibinfo  {journal} {Nature Materials}\ }\textbf {\bibinfo {volume} {19}},\ \bibinfo {pages} {830} (\bibinfo {year} {2020})}\BibitemShut {NoStop}%
\bibitem [{\citenamefont {Morimoto}\ \emph {et~al.}(2023)\citenamefont {Morimoto}, \citenamefont {Kitamura},\ and\ \citenamefont {Nagaosa}}]{morimoto2023review}%
  \BibitemOpen
  \bibfield  {author} {\bibinfo {author} {\bibfnamefont {T.}~\bibnamefont {Morimoto}}, \bibinfo {author} {\bibfnamefont {S.}~\bibnamefont {Kitamura}},\ and\ \bibinfo {author} {\bibfnamefont {N.}~\bibnamefont {Nagaosa}},\ }\bibfield  {title} {\bibinfo {title} {Geometric aspects of nonlinear and nonequilibrium phenomena},\ }\href {https://doi.org/10.7566/JPSJ.92.072001} {\bibfield  {journal} {\bibinfo  {journal} {Journal of the Physical Society of Japan}\ }\textbf {\bibinfo {volume} {92}},\ \bibinfo {pages} {072001} (\bibinfo {year} {2023})}\BibitemShut {NoStop}%
\bibitem [{\citenamefont {Morimoto}\ \emph {et~al.}(2016)\citenamefont {Morimoto}, \citenamefont {Zhong}, \citenamefont {Orenstein},\ and\ \citenamefont {Moore}}]{Morimoto_b2016}%
  \BibitemOpen
  \bibfield  {author} {\bibinfo {author} {\bibfnamefont {T.}~\bibnamefont {Morimoto}}, \bibinfo {author} {\bibfnamefont {S.}~\bibnamefont {Zhong}}, \bibinfo {author} {\bibfnamefont {J.}~\bibnamefont {Orenstein}},\ and\ \bibinfo {author} {\bibfnamefont {J.~E.}\ \bibnamefont {Moore}},\ }\bibfield  {title} {\bibinfo {title} {Semiclassical theory of nonlinear magneto-optical responses with applications to topological dirac/weyl semimetals},\ }\href {https://doi.org/10.1103/PhysRevB.94.245121} {\bibfield  {journal} {\bibinfo  {journal} {Physical Review B}\ }\textbf {\bibinfo {volume} {94}},\ \bibinfo {pages} {245121} (\bibinfo {year} {2016})}\BibitemShut {NoStop}%
\bibitem [{\citenamefont {Chan}\ \emph {et~al.}(2017)\citenamefont {Chan}, \citenamefont {Lindner}, \citenamefont {Refael},\ and\ \citenamefont {Lee}}]{Chan2017}%
  \BibitemOpen
  \bibfield  {author} {\bibinfo {author} {\bibfnamefont {C.-K.}\ \bibnamefont {Chan}}, \bibinfo {author} {\bibfnamefont {N.~H.}\ \bibnamefont {Lindner}}, \bibinfo {author} {\bibfnamefont {G.}~\bibnamefont {Refael}},\ and\ \bibinfo {author} {\bibfnamefont {P.~A.}\ \bibnamefont {Lee}},\ }\bibfield  {title} {\bibinfo {title} {Photocurrents in weyl semimetals},\ }\href {https://doi.org/10.1103/PhysRevB.95.041104} {\bibfield  {journal} {\bibinfo  {journal} {Physical Review B}\ }\textbf {\bibinfo {volume} {95}},\ \bibinfo {pages} {041104} (\bibinfo {year} {2017})}\BibitemShut {NoStop}%
\bibitem [{\citenamefont {Yang}\ \emph {et~al.}(2018)\citenamefont {Yang}, \citenamefont {Burch},\ and\ \citenamefont {Ran}}]{Yang2018}%
  \BibitemOpen
  \bibfield  {author} {\bibinfo {author} {\bibfnamefont {X.}~\bibnamefont {Yang}}, \bibinfo {author} {\bibfnamefont {K.}~\bibnamefont {Burch}},\ and\ \bibinfo {author} {\bibfnamefont {Y.}~\bibnamefont {Ran}},\ }\href {https://arxiv.org/abs/1712.09363} {\bibinfo {title} {Divergent bulk photovoltaic effect in weyl semimetals}} (\bibinfo {year} {2018}),\ \Eprint {https://arxiv.org/abs/1712.09363} {arXiv:1712.09363 [cond-mat.mes-hall]} \BibitemShut {NoStop}%
\bibitem [{\citenamefont {Parker}\ \emph {et~al.}(2019)\citenamefont {Parker}, \citenamefont {Morimoto}, \citenamefont {Orenstein},\ and\ \citenamefont {Moore}}]{Parker2019}%
  \BibitemOpen
  \bibfield  {author} {\bibinfo {author} {\bibfnamefont {D.~E.}\ \bibnamefont {Parker}}, \bibinfo {author} {\bibfnamefont {T.}~\bibnamefont {Morimoto}}, \bibinfo {author} {\bibfnamefont {J.}~\bibnamefont {Orenstein}},\ and\ \bibinfo {author} {\bibfnamefont {J.~E.}\ \bibnamefont {Moore}},\ }\bibfield  {title} {\bibinfo {title} {Diagrammatic approach to nonlinear optical response with application to weyl semimetals},\ }\href {https://doi.org/10.1103/PhysRevB.99.045121} {\bibfield  {journal} {\bibinfo  {journal} {Physical Review B}\ }\textbf {\bibinfo {volume} {99}},\ \bibinfo {pages} {045121} (\bibinfo {year} {2019})}\BibitemShut {NoStop}%
\bibitem [{\citenamefont {Avdoshkin}\ \emph {et~al.}(2020)\citenamefont {Avdoshkin}, \citenamefont {Kozii},\ and\ \citenamefont {Moore}}]{Avdoshkin2020}%
  \BibitemOpen
  \bibfield  {author} {\bibinfo {author} {\bibfnamefont {A.}~\bibnamefont {Avdoshkin}}, \bibinfo {author} {\bibfnamefont {V.}~\bibnamefont {Kozii}},\ and\ \bibinfo {author} {\bibfnamefont {J.~E.}\ \bibnamefont {Moore}},\ }\bibfield  {title} {\bibinfo {title} {Interactions remove the quantization of the chiral photocurrent at weyl points},\ }\href {https://doi.org/10.1103/PhysRevLett.124.196603} {\bibfield  {journal} {\bibinfo  {journal} {Physical Review Letter}\ }\textbf {\bibinfo {volume} {124}},\ \bibinfo {pages} {196603} (\bibinfo {year} {2020})}\BibitemShut {NoStop}%
\bibitem [{\citenamefont {Ahn}\ \emph {et~al.}(2020)\citenamefont {Ahn}, \citenamefont {Guo},\ and\ \citenamefont {Nagaosa}}]{Ahn2020}%
  \BibitemOpen
  \bibfield  {author} {\bibinfo {author} {\bibfnamefont {J.}~\bibnamefont {Ahn}}, \bibinfo {author} {\bibfnamefont {G.-Y.}\ \bibnamefont {Guo}},\ and\ \bibinfo {author} {\bibfnamefont {N.}~\bibnamefont {Nagaosa}},\ }\bibfield  {title} {\bibinfo {title} {Low-frequency divergence and quantum geometry of the bulk photovoltaic effect in topological semimetals},\ }\href {https://doi.org/10.1103/PhysRevX.10.041041} {\bibfield  {journal} {\bibinfo  {journal} {Physical Review X}\ }\textbf {\bibinfo {volume} {10}},\ \bibinfo {pages} {041041} (\bibinfo {year} {2020})}\BibitemShut {NoStop}%
\bibitem [{\citenamefont {Watanabe}\ and\ \citenamefont {Yanase}(2021)}]{Watanabe2021}%
  \BibitemOpen
  \bibfield  {author} {\bibinfo {author} {\bibfnamefont {H.}~\bibnamefont {Watanabe}}\ and\ \bibinfo {author} {\bibfnamefont {Y.}~\bibnamefont {Yanase}},\ }\bibfield  {title} {\bibinfo {title} {Chiral photocurrent in parity-violating magnet and enhanced response in topological antiferromagnet},\ }\href {https://doi.org/10.1103/PhysRevX.11.011001} {\bibfield  {journal} {\bibinfo  {journal} {Physical Review X}\ }\textbf {\bibinfo {volume} {11}},\ \bibinfo {pages} {011001} (\bibinfo {year} {2021})}\BibitemShut {NoStop}%
\bibitem [{\citenamefont {Takasan}\ \emph {et~al.}(2021)\citenamefont {Takasan}, \citenamefont {Morimoto}, \citenamefont {Orenstein},\ and\ \citenamefont {Moore}}]{Takasan2021}%
  \BibitemOpen
  \bibfield  {author} {\bibinfo {author} {\bibfnamefont {K.}~\bibnamefont {Takasan}}, \bibinfo {author} {\bibfnamefont {T.}~\bibnamefont {Morimoto}}, \bibinfo {author} {\bibfnamefont {J.}~\bibnamefont {Orenstein}},\ and\ \bibinfo {author} {\bibfnamefont {J.~E.}\ \bibnamefont {Moore}},\ }\bibfield  {title} {\bibinfo {title} {Current-induced second harmonic generation in inversion-symmetric dirac and weyl semimetals},\ }\href {https://doi.org/10.1103/PhysRevB.104.L161202} {\bibfield  {journal} {\bibinfo  {journal} {Physical Review B}\ }\textbf {\bibinfo {volume} {104}},\ \bibinfo {pages} {L161202} (\bibinfo {year} {2021})}\BibitemShut {NoStop}%
\bibitem [{\citenamefont {Ikeda}\ \emph {et~al.}(2023)\citenamefont {Ikeda}, \citenamefont {Kitamura},\ and\ \citenamefont {Morimoto}}]{Ikeda2023}%
  \BibitemOpen
  \bibfield  {author} {\bibinfo {author} {\bibfnamefont {Y.}~\bibnamefont {Ikeda}}, \bibinfo {author} {\bibfnamefont {S.}~\bibnamefont {Kitamura}},\ and\ \bibinfo {author} {\bibfnamefont {T.}~\bibnamefont {Morimoto}},\ }\bibfield  {title} {\bibinfo {title} {Photocurrent induced by a bicircular light drive in centrosymmetric systems},\ }\href {https://doi.org/10.1103/PhysRevLett.131.096301} {\bibfield  {journal} {\bibinfo  {journal} {Physical Review Letters}\ }\textbf {\bibinfo {volume} {131}},\ \bibinfo {pages} {096301} (\bibinfo {year} {2023})}\BibitemShut {NoStop}%
\bibitem [{\citenamefont {Ahn}(2023)}]{Ahn2023}%
  \BibitemOpen
  \bibfield  {author} {\bibinfo {author} {\bibfnamefont {J.}~\bibnamefont {Ahn}},\ }\bibfield  {title} {\bibinfo {title} {Topological enhancement of nonlinear transport in unconventional point-node semimetals},\ }\href {https://doi.org/10.1103/PhysRevB.107.L201112} {\bibfield  {journal} {\bibinfo  {journal} {Physical Review B}\ }\textbf {\bibinfo {volume} {107}},\ \bibinfo {pages} {L201112} (\bibinfo {year} {2023})}\BibitemShut {NoStop}%
\bibitem [{\citenamefont {Hsu}\ \emph {et~al.}(2023)\citenamefont {Hsu}, \citenamefont {You}, \citenamefont {Ahn},\ and\ \citenamefont {Guo}}]{Hsu2023}%
  \BibitemOpen
  \bibfield  {author} {\bibinfo {author} {\bibfnamefont {H.-C.}\ \bibnamefont {Hsu}}, \bibinfo {author} {\bibfnamefont {J.-S.}\ \bibnamefont {You}}, \bibinfo {author} {\bibfnamefont {J.}~\bibnamefont {Ahn}},\ and\ \bibinfo {author} {\bibfnamefont {G.-Y.}\ \bibnamefont {Guo}},\ }\bibfield  {title} {\bibinfo {title} {Nonlinear photoconductivities and quantum geometry of chiral multifold fermions},\ }\href {https://doi.org/10.1103/PhysRevB.107.155434} {\bibfield  {journal} {\bibinfo  {journal} {Physical Review B}\ }\textbf {\bibinfo {volume} {107}},\ \bibinfo {pages} {155434} (\bibinfo {year} {2023})}\BibitemShut {NoStop}%
\bibitem [{\citenamefont {Ikeda}\ \emph {et~al.}(2024)\citenamefont {Ikeda}, \citenamefont {Kitamura},\ and\ \citenamefont {Morimoto}}]{Ikeda2024}%
  \BibitemOpen
  \bibfield  {author} {\bibinfo {author} {\bibfnamefont {Y.}~\bibnamefont {Ikeda}}, \bibinfo {author} {\bibfnamefont {S.}~\bibnamefont {Kitamura}},\ and\ \bibinfo {author} {\bibfnamefont {T.}~\bibnamefont {Morimoto}},\ }\bibfield  {title} {\bibinfo {title} {Controllable photocurrent generation in dirac systems with two frequency drives},\ }\href {https://doi.org/10.1103/PhysRevB.110.235206} {\bibfield  {journal} {\bibinfo  {journal} {Physical Review B}\ }\textbf {\bibinfo {volume} {110}},\ \bibinfo {pages} {235206} (\bibinfo {year} {2024})}\BibitemShut {NoStop}%
\bibitem [{\citenamefont {Yoshida}\ and\ \citenamefont {Murakami}(2025)}]{Yoshida2025}%
  \BibitemOpen
  \bibfield  {author} {\bibinfo {author} {\bibfnamefont {H.}~\bibnamefont {Yoshida}}\ and\ \bibinfo {author} {\bibfnamefont {S.}~\bibnamefont {Murakami}},\ }\bibfield  {title} {\bibinfo {title} {Diverging shift current responses in the gapless limit of two-dimensional systems},\ }\href {https://doi.org/10.1103/PhysRevB.111.155402} {\bibfield  {journal} {\bibinfo  {journal} {Physical Review B}\ }\textbf {\bibinfo {volume} {111}},\ \bibinfo {pages} {155402} (\bibinfo {year} {2025})}\BibitemShut {NoStop}%
\bibitem [{\citenamefont {Mei}\ and\ \citenamefont {Liu}(2025)}]{Mei2025}%
  \BibitemOpen
  \bibfield  {author} {\bibinfo {author} {\bibfnamefont {R.}~\bibnamefont {Mei}}\ and\ \bibinfo {author} {\bibfnamefont {C.-X.}\ \bibnamefont {Liu}},\ }\bibfield  {title} {\bibinfo {title} {Magnetic-resonance-induced non-linear current response in magnetic weyl semimetals},\ }\href {https://doi.org/10.1007/s43673-025-00145-x} {\bibfield  {journal} {\bibinfo  {journal} {AAPPS Bulletin}\ }\textbf {\bibinfo {volume} {35}},\ \bibinfo {pages} {4} (\bibinfo {year} {2025})}\BibitemShut {NoStop}%
\bibitem [{\citenamefont {Wu}\ \emph {et~al.}(2017)\citenamefont {Wu}, \citenamefont {Patankar}, \citenamefont {Morimoto}, \citenamefont {Nair}, \citenamefont {Thewalt}, \citenamefont {Little}, \citenamefont {Analytis}, \citenamefont {Moore},\ and\ \citenamefont {Orenstein}}]{Wu2017}%
  \BibitemOpen
  \bibfield  {author} {\bibinfo {author} {\bibfnamefont {L.}~\bibnamefont {Wu}}, \bibinfo {author} {\bibfnamefont {S.}~\bibnamefont {Patankar}}, \bibinfo {author} {\bibfnamefont {T.}~\bibnamefont {Morimoto}}, \bibinfo {author} {\bibfnamefont {N.~L.}\ \bibnamefont {Nair}}, \bibinfo {author} {\bibfnamefont {E.}~\bibnamefont {Thewalt}}, \bibinfo {author} {\bibfnamefont {A.}~\bibnamefont {Little}}, \bibinfo {author} {\bibfnamefont {J.~G.}\ \bibnamefont {Analytis}}, \bibinfo {author} {\bibfnamefont {J.~E.}\ \bibnamefont {Moore}},\ and\ \bibinfo {author} {\bibfnamefont {J.}~\bibnamefont {Orenstein}},\ }\bibfield  {title} {\bibinfo {title} {Giant anisotropic nonlinear optical response in transition metal monopnictide weyl semimetals},\ }\href {https://doi.org/10.1038/nphys3969} {\bibfield  {journal} {\bibinfo  {journal} {Nature Physics}\ }\textbf {\bibinfo {volume} {13}},\ \bibinfo {pages} {350} (\bibinfo {year} {2017})}\BibitemShut {NoStop}%
\bibitem [{\citenamefont {Ma}\ \emph {et~al.}(2019)\citenamefont {Ma}, \citenamefont {Gu}, \citenamefont {Liu}, \citenamefont {Lai}, \citenamefont {Yu}, \citenamefont {Zhuo}, \citenamefont {Liu}, \citenamefont {Chen}, \citenamefont {Feng},\ and\ \citenamefont {Sun}}]{Ma2019}%
  \BibitemOpen
  \bibfield  {author} {\bibinfo {author} {\bibfnamefont {J.}~\bibnamefont {Ma}}, \bibinfo {author} {\bibfnamefont {Q.}~\bibnamefont {Gu}}, \bibinfo {author} {\bibfnamefont {Y.}~\bibnamefont {Liu}}, \bibinfo {author} {\bibfnamefont {J.}~\bibnamefont {Lai}}, \bibinfo {author} {\bibfnamefont {P.}~\bibnamefont {Yu}}, \bibinfo {author} {\bibfnamefont {X.}~\bibnamefont {Zhuo}}, \bibinfo {author} {\bibfnamefont {Z.}~\bibnamefont {Liu}}, \bibinfo {author} {\bibfnamefont {J.-H.}\ \bibnamefont {Chen}}, \bibinfo {author} {\bibfnamefont {J.}~\bibnamefont {Feng}},\ and\ \bibinfo {author} {\bibfnamefont {D.}~\bibnamefont {Sun}},\ }\bibfield  {title} {\bibinfo {title} {Nonlinear photoresponse of type-ii weyl semimetals},\ }\href {https://doi.org/10.1038/s41563-019-0296-5} {\bibfield  {journal} {\bibinfo  {journal} {Nature Materials}\ }\textbf {\bibinfo {volume} {18}},\ \bibinfo {pages} {476} (\bibinfo {year} {2019})}\BibitemShut {NoStop}%
\bibitem [{\citenamefont {Osterhoudt}\ \emph {et~al.}(2019)\citenamefont {Osterhoudt}, \citenamefont {Diebel}, \citenamefont {Gray}, \citenamefont {Yang}, \citenamefont {Stanco}, \citenamefont {Huang}, \citenamefont {Shen}, \citenamefont {Ni}, \citenamefont {Moll}, \citenamefont {Ran},\ and\ \citenamefont {Burch}}]{Osterhoudt2019}%
  \BibitemOpen
  \bibfield  {author} {\bibinfo {author} {\bibfnamefont {G.~B.}\ \bibnamefont {Osterhoudt}}, \bibinfo {author} {\bibfnamefont {L.~K.}\ \bibnamefont {Diebel}}, \bibinfo {author} {\bibfnamefont {M.~J.}\ \bibnamefont {Gray}}, \bibinfo {author} {\bibfnamefont {X.}~\bibnamefont {Yang}}, \bibinfo {author} {\bibfnamefont {J.}~\bibnamefont {Stanco}}, \bibinfo {author} {\bibfnamefont {X.}~\bibnamefont {Huang}}, \bibinfo {author} {\bibfnamefont {B.}~\bibnamefont {Shen}}, \bibinfo {author} {\bibfnamefont {N.}~\bibnamefont {Ni}}, \bibinfo {author} {\bibfnamefont {P.~J.~W.}\ \bibnamefont {Moll}}, \bibinfo {author} {\bibfnamefont {Y.}~\bibnamefont {Ran}},\ and\ \bibinfo {author} {\bibfnamefont {K.~S.}\ \bibnamefont {Burch}},\ }\bibfield  {title} {\bibinfo {title} {Colossal mid-infrared bulk photovoltaic effect in a type-i weyl semimetal},\ }\href {https://doi.org/10.1038/s41563-019-0297-4} {\bibfield  {journal} {\bibinfo  {journal} {Nature Materials}\ }\textbf {\bibinfo {volume} {18}},\ \bibinfo {pages} {471} (\bibinfo
  {year} {2019})}\BibitemShut {NoStop}%
\bibitem [{\citenamefont {Murakami}(2007)}]{Murakami2007}%
  \BibitemOpen
  \bibfield  {author} {\bibinfo {author} {\bibfnamefont {S.}~\bibnamefont {Murakami}},\ }\bibfield  {title} {\bibinfo {title} {Phase transition between the quantum spin hall and insulator phases in 3d: emergence of a topological gapless phase},\ }\href {https://doi.org/10.1088/1367-2630/9/9/356} {\bibfield  {journal} {\bibinfo  {journal} {New Journal of Physics}\ }\textbf {\bibinfo {volume} {9}},\ \bibinfo {pages} {356} (\bibinfo {year} {2007})}\BibitemShut {NoStop}%
\bibitem [{\citenamefont {Wan}\ \emph {et~al.}(2011)\citenamefont {Wan}, \citenamefont {Turner}, \citenamefont {Vishwanath},\ and\ \citenamefont {Savrasov}}]{Wan2011}%
  \BibitemOpen
  \bibfield  {author} {\bibinfo {author} {\bibfnamefont {X.}~\bibnamefont {Wan}}, \bibinfo {author} {\bibfnamefont {A.~M.}\ \bibnamefont {Turner}}, \bibinfo {author} {\bibfnamefont {A.}~\bibnamefont {Vishwanath}},\ and\ \bibinfo {author} {\bibfnamefont {S.~Y.}\ \bibnamefont {Savrasov}},\ }\bibfield  {title} {\bibinfo {title} {Topological semimetal and fermi-arc surface states in the electronic structure of pyrochlore iridates},\ }\href {https://doi.org/10.1103/PhysRevB.83.205101} {\bibfield  {journal} {\bibinfo  {journal} {Physical Review B}\ }\textbf {\bibinfo {volume} {83}},\ \bibinfo {pages} {205101} (\bibinfo {year} {2011})}\BibitemShut {NoStop}%
\bibitem [{\citenamefont {Burkov}\ and\ \citenamefont {Balents}(2011)}]{Burkov2011}%
  \BibitemOpen
  \bibfield  {author} {\bibinfo {author} {\bibfnamefont {A.~A.}\ \bibnamefont {Burkov}}\ and\ \bibinfo {author} {\bibfnamefont {L.}~\bibnamefont {Balents}},\ }\bibfield  {title} {\bibinfo {title} {Weyl semimetal in a topological insulator multilayer},\ }\href {https://doi.org/10.1103/PhysRevLett.107.127205} {\bibfield  {journal} {\bibinfo  {journal} {Physical Review Letter}\ }\textbf {\bibinfo {volume} {107}},\ \bibinfo {pages} {127205} (\bibinfo {year} {2011})}\BibitemShut {NoStop}%
\bibitem [{\citenamefont {Xu}\ \emph {et~al.}(2015)\citenamefont {Xu}, \citenamefont {Belopolski}, \citenamefont {Alidoust}, \citenamefont {Neupane}, \citenamefont {Bian}, \citenamefont {Zhang}, \citenamefont {Sankar}, \citenamefont {Chang}, \citenamefont {Yuan}, \citenamefont {Lee}, \citenamefont {Huang}, \citenamefont {Zheng}, \citenamefont {Ma}, \citenamefont {Sanchez}, \citenamefont {Wang}, \citenamefont {Bansil}, \citenamefont {Chou}, \citenamefont {Shibayev}, \citenamefont {Lin}, \citenamefont {Jia},\ and\ \citenamefont {Hasan}}]{Xu2015}%
  \BibitemOpen
  \bibfield  {author} {\bibinfo {author} {\bibfnamefont {S.-Y.}\ \bibnamefont {Xu}}, \bibinfo {author} {\bibfnamefont {I.}~\bibnamefont {Belopolski}}, \bibinfo {author} {\bibfnamefont {N.}~\bibnamefont {Alidoust}}, \bibinfo {author} {\bibfnamefont {M.}~\bibnamefont {Neupane}}, \bibinfo {author} {\bibfnamefont {G.}~\bibnamefont {Bian}}, \bibinfo {author} {\bibfnamefont {C.}~\bibnamefont {Zhang}}, \bibinfo {author} {\bibfnamefont {R.}~\bibnamefont {Sankar}}, \bibinfo {author} {\bibfnamefont {G.}~\bibnamefont {Chang}}, \bibinfo {author} {\bibfnamefont {Z.}~\bibnamefont {Yuan}}, \bibinfo {author} {\bibfnamefont {C.-C.}\ \bibnamefont {Lee}}, \bibinfo {author} {\bibfnamefont {S.-M.}\ \bibnamefont {Huang}}, \bibinfo {author} {\bibfnamefont {H.}~\bibnamefont {Zheng}}, \bibinfo {author} {\bibfnamefont {J.}~\bibnamefont {Ma}}, \bibinfo {author} {\bibfnamefont {D.~S.}\ \bibnamefont {Sanchez}}, \bibinfo {author} {\bibfnamefont {B.}~\bibnamefont {Wang}}, \bibinfo {author} {\bibfnamefont {A.}~\bibnamefont {Bansil}},
  \bibinfo {author} {\bibfnamefont {F.}~\bibnamefont {Chou}}, \bibinfo {author} {\bibfnamefont {P.~P.}\ \bibnamefont {Shibayev}}, \bibinfo {author} {\bibfnamefont {H.}~\bibnamefont {Lin}}, \bibinfo {author} {\bibfnamefont {S.}~\bibnamefont {Jia}},\ and\ \bibinfo {author} {\bibfnamefont {M.~Z.}\ \bibnamefont {Hasan}},\ }\bibfield  {title} {\bibinfo {title} {Discovery of a weyl fermion semimetal and topological fermi arcs},\ }\href {https://doi.org/10.1126/science.aaa9297} {\bibfield  {journal} {\bibinfo  {journal} {Science}\ }\textbf {\bibinfo {volume} {349}},\ \bibinfo {pages} {613} (\bibinfo {year} {2015})}\BibitemShut {NoStop}%
\bibitem [{\citenamefont {de~Juan}\ \emph {et~al.}(2017)\citenamefont {de~Juan}, \citenamefont {Grushin}, \citenamefont {Morimoto},\ and\ \citenamefont {Moore}}]{deJuan2017}%
  \BibitemOpen
  \bibfield  {author} {\bibinfo {author} {\bibfnamefont {F.}~\bibnamefont {de~Juan}}, \bibinfo {author} {\bibfnamefont {A.~G.}\ \bibnamefont {Grushin}}, \bibinfo {author} {\bibfnamefont {T.}~\bibnamefont {Morimoto}},\ and\ \bibinfo {author} {\bibfnamefont {J.~E.}\ \bibnamefont {Moore}},\ }\bibfield  {title} {\bibinfo {title} {Quantized circular photogalvanic effect in weyl semimetals},\ }\href {https://doi.org/10.1038/ncomms15995} {\bibfield  {journal} {\bibinfo  {journal} {Nature Communications}\ }\textbf {\bibinfo {volume} {8}},\ \bibinfo {pages} {15995} (\bibinfo {year} {2017})}\BibitemShut {NoStop}%
\bibitem [{\citenamefont {Flicker}\ \emph {et~al.}(2018)\citenamefont {Flicker}, \citenamefont {de~Juan}, \citenamefont {Bradlyn}, \citenamefont {Morimoto}, \citenamefont {Vergniory},\ and\ \citenamefont {Grushin}}]{Felix2018}%
  \BibitemOpen
  \bibfield  {author} {\bibinfo {author} {\bibfnamefont {F.}~\bibnamefont {Flicker}}, \bibinfo {author} {\bibfnamefont {F.}~\bibnamefont {de~Juan}}, \bibinfo {author} {\bibfnamefont {B.}~\bibnamefont {Bradlyn}}, \bibinfo {author} {\bibfnamefont {T.}~\bibnamefont {Morimoto}}, \bibinfo {author} {\bibfnamefont {M.~G.}\ \bibnamefont {Vergniory}},\ and\ \bibinfo {author} {\bibfnamefont {A.~G.}\ \bibnamefont {Grushin}},\ }\bibfield  {title} {\bibinfo {title} {Chiral optical response of multifold fermions},\ }\href {https://doi.org/10.1103/PhysRevB.98.155145} {\bibfield  {journal} {\bibinfo  {journal} {Physical Review B}\ }\textbf {\bibinfo {volume} {98}},\ \bibinfo {pages} {155145} (\bibinfo {year} {2018})}\BibitemShut {NoStop}%
\bibitem [{\citenamefont {de~Juan}\ \emph {et~al.}(2020)\citenamefont {de~Juan}, \citenamefont {Zhang}, \citenamefont {Morimoto}, \citenamefont {Sun}, \citenamefont {Moore},\ and\ \citenamefont {Grushin}}]{deJuan2020}%
  \BibitemOpen
  \bibfield  {author} {\bibinfo {author} {\bibfnamefont {F.}~\bibnamefont {de~Juan}}, \bibinfo {author} {\bibfnamefont {Y.}~\bibnamefont {Zhang}}, \bibinfo {author} {\bibfnamefont {T.}~\bibnamefont {Morimoto}}, \bibinfo {author} {\bibfnamefont {Y.}~\bibnamefont {Sun}}, \bibinfo {author} {\bibfnamefont {J.~E.}\ \bibnamefont {Moore}},\ and\ \bibinfo {author} {\bibfnamefont {A.~G.}\ \bibnamefont {Grushin}},\ }\bibfield  {title} {\bibinfo {title} {Difference frequency generation in topological semimetals},\ }\href {https://doi.org/10.1103/PhysRevResearch.2.012017} {\bibfield  {journal} {\bibinfo  {journal} {Physical Review Research}\ }\textbf {\bibinfo {volume} {2}},\ \bibinfo {pages} {012017} (\bibinfo {year} {2020})}\BibitemShut {NoStop}%
\bibitem [{\citenamefont {Le}\ \emph {et~al.}(2020)\citenamefont {Le}, \citenamefont {Zhang}, \citenamefont {Felser},\ and\ \citenamefont {Sun}}]{Le2020}%
  \BibitemOpen
  \bibfield  {author} {\bibinfo {author} {\bibfnamefont {C.}~\bibnamefont {Le}}, \bibinfo {author} {\bibfnamefont {Y.}~\bibnamefont {Zhang}}, \bibinfo {author} {\bibfnamefont {C.}~\bibnamefont {Felser}},\ and\ \bibinfo {author} {\bibfnamefont {Y.}~\bibnamefont {Sun}},\ }\bibfield  {title} {\bibinfo {title} {Ab initio study of quantized circular photogalvanic effect in chiral multifold semimetals},\ }\href {https://doi.org/10.1103/PhysRevB.102.121111} {\bibfield  {journal} {\bibinfo  {journal} {Physical Review B}\ }\textbf {\bibinfo {volume} {102}},\ \bibinfo {pages} {121111} (\bibinfo {year} {2020})}\BibitemShut {NoStop}%
\bibitem [{\citenamefont {Raj}\ \emph {et~al.}(2024)\citenamefont {Raj}, \citenamefont {Chaudhary},\ and\ \citenamefont {Fiete}}]{Raj2024}%
  \BibitemOpen
  \bibfield  {author} {\bibinfo {author} {\bibfnamefont {A.}~\bibnamefont {Raj}}, \bibinfo {author} {\bibfnamefont {S.}~\bibnamefont {Chaudhary}},\ and\ \bibinfo {author} {\bibfnamefont {G.~A.}\ \bibnamefont {Fiete}},\ }\bibfield  {title} {\bibinfo {title} {Photogalvanic response in multi-weyl semimetals},\ }\href {https://doi.org/10.1103/PhysRevResearch.6.013048} {\bibfield  {journal} {\bibinfo  {journal} {Physical Review Research}\ }\textbf {\bibinfo {volume} {6}},\ \bibinfo {pages} {013048} (\bibinfo {year} {2024})}\BibitemShut {NoStop}%
\bibitem [{\citenamefont {Cao}\ \emph {et~al.}(2024)\citenamefont {Cao}, \citenamefont {Zeng}, \citenamefont {Li}, \citenamefont {Wang}, \citenamefont {Yang}, \citenamefont {Yu},\ and\ \citenamefont {Yao}}]{Cao2024}%
  \BibitemOpen
  \bibfield  {author} {\bibinfo {author} {\bibfnamefont {J.}~\bibnamefont {Cao}}, \bibinfo {author} {\bibfnamefont {C.}~\bibnamefont {Zeng}}, \bibinfo {author} {\bibfnamefont {X.-P.}\ \bibnamefont {Li}}, \bibinfo {author} {\bibfnamefont {M.}~\bibnamefont {Wang}}, \bibinfo {author} {\bibfnamefont {S.~A.}\ \bibnamefont {Yang}}, \bibinfo {author} {\bibfnamefont {Z.-M.}\ \bibnamefont {Yu}},\ and\ \bibinfo {author} {\bibfnamefont {Y.}~\bibnamefont {Yao}},\ }\bibfield  {title} {\bibinfo {title} {Low-frequency divergence of circular photomagnetic effect in topological semimetals},\ }\href {https://doi.org/10.1103/PhysRevB.110.L041114} {\bibfield  {journal} {\bibinfo  {journal} {Physical Review B}\ }\textbf {\bibinfo {volume} {110}},\ \bibinfo {pages} {L041114} (\bibinfo {year} {2024})}\BibitemShut {NoStop}%
\bibitem [{\citenamefont {Ahn}\ and\ \citenamefont {Ghosh}(2023)}]{Ahn2024}%
  \BibitemOpen
  \bibfield  {author} {\bibinfo {author} {\bibfnamefont {J.}~\bibnamefont {Ahn}}\ and\ \bibinfo {author} {\bibfnamefont {B.}~\bibnamefont {Ghosh}},\ }\bibfield  {title} {\bibinfo {title} {Topological circular dichroism in chiral multifold semimetals},\ }\href {https://doi.org/10.1103/PhysRevLett.131.116603} {\bibfield  {journal} {\bibinfo  {journal} {Physical Review Letters}\ }\textbf {\bibinfo {volume} {131}},\ \bibinfo {pages} {116603} (\bibinfo {year} {2023})}\BibitemShut {NoStop}%
\bibitem [{\citenamefont {Rees}\ \emph {et~al.}(2020)\citenamefont {Rees}, \citenamefont {Manna}, \citenamefont {Lu}, \citenamefont {Morimoto}, \citenamefont {Borrmann}, \citenamefont {Felser}, \citenamefont {Moore}, \citenamefont {Torchinsky},\ and\ \citenamefont {Orenstein}}]{Rees2020}%
  \BibitemOpen
  \bibfield  {author} {\bibinfo {author} {\bibfnamefont {D.}~\bibnamefont {Rees}}, \bibinfo {author} {\bibfnamefont {K.}~\bibnamefont {Manna}}, \bibinfo {author} {\bibfnamefont {B.}~\bibnamefont {Lu}}, \bibinfo {author} {\bibfnamefont {T.}~\bibnamefont {Morimoto}}, \bibinfo {author} {\bibfnamefont {H.}~\bibnamefont {Borrmann}}, \bibinfo {author} {\bibfnamefont {C.}~\bibnamefont {Felser}}, \bibinfo {author} {\bibfnamefont {J.~E.}\ \bibnamefont {Moore}}, \bibinfo {author} {\bibfnamefont {D.~H.}\ \bibnamefont {Torchinsky}},\ and\ \bibinfo {author} {\bibfnamefont {J.}~\bibnamefont {Orenstein}},\ }\bibfield  {title} {\bibinfo {title} {Helicity-dependent photocurrents in the chiral weyl semimetal rhsi},\ }\href {https://doi.org/10.1126/sciadv.aba0509} {\bibfield  {journal} {\bibinfo  {journal} {Science Advances}\ }\textbf {\bibinfo {volume} {6}},\ \bibinfo {pages} {eaba0509} (\bibinfo {year} {2020})}\BibitemShut {NoStop}%
\bibitem [{\citenamefont {Ni}\ \emph {et~al.}(2021)\citenamefont {Ni}, \citenamefont {Wang}, \citenamefont {Zhang}, \citenamefont {Pozo}, \citenamefont {Xu}, \citenamefont {Han}, \citenamefont {Manna}, \citenamefont {Paglione}, \citenamefont {Felser}, \citenamefont {Grushin}, \citenamefont {de~Juan}, \citenamefont {Mele},\ and\ \citenamefont {Wu}}]{Ni2021}%
  \BibitemOpen
  \bibfield  {author} {\bibinfo {author} {\bibfnamefont {Z.}~\bibnamefont {Ni}}, \bibinfo {author} {\bibfnamefont {K.}~\bibnamefont {Wang}}, \bibinfo {author} {\bibfnamefont {Y.}~\bibnamefont {Zhang}}, \bibinfo {author} {\bibfnamefont {O.}~\bibnamefont {Pozo}}, \bibinfo {author} {\bibfnamefont {B.}~\bibnamefont {Xu}}, \bibinfo {author} {\bibfnamefont {X.}~\bibnamefont {Han}}, \bibinfo {author} {\bibfnamefont {K.}~\bibnamefont {Manna}}, \bibinfo {author} {\bibfnamefont {J.}~\bibnamefont {Paglione}}, \bibinfo {author} {\bibfnamefont {C.}~\bibnamefont {Felser}}, \bibinfo {author} {\bibfnamefont {A.~G.}\ \bibnamefont {Grushin}}, \bibinfo {author} {\bibfnamefont {F.}~\bibnamefont {de~Juan}}, \bibinfo {author} {\bibfnamefont {E.~J.}\ \bibnamefont {Mele}},\ and\ \bibinfo {author} {\bibfnamefont {L.}~\bibnamefont {Wu}},\ }\bibfield  {title} {\bibinfo {title} {Giant topological longitudinal circular photo-galvanic effect in the chiral multifold semimetal cosi},\ }\href {https://doi.org/10.1038/s41467-020-20408-5}
  {\bibfield  {journal} {\bibinfo  {journal} {Nature Communications}\ }\textbf {\bibinfo {volume} {12}},\ \bibinfo {pages} {154} (\bibinfo {year} {2021})}\BibitemShut {NoStop}%
\bibitem [{\citenamefont {Morimoto}\ and\ \citenamefont {Nagaosa}(2016)}]{Morimoto2016_exciton}%
  \BibitemOpen
  \bibfield  {author} {\bibinfo {author} {\bibfnamefont {T.}~\bibnamefont {Morimoto}}\ and\ \bibinfo {author} {\bibfnamefont {N.}~\bibnamefont {Nagaosa}},\ }\bibfield  {title} {\bibinfo {title} {Topological aspects of nonlinear excitonic processes in noncentrosymmetric crystals},\ }\href {https://doi.org/10.1103/PhysRevB.94.035117} {\bibfield  {journal} {\bibinfo  {journal} {Physical Review B}\ }\textbf {\bibinfo {volume} {94}},\ \bibinfo {pages} {035117} (\bibinfo {year} {2016})}\BibitemShut {NoStop}%
\bibitem [{\citenamefont {Morimoto}\ and\ \citenamefont {Nagaosa}(2019)}]{Morimoto2019_electromagnon}%
  \BibitemOpen
  \bibfield  {author} {\bibinfo {author} {\bibfnamefont {T.}~\bibnamefont {Morimoto}}\ and\ \bibinfo {author} {\bibfnamefont {N.}~\bibnamefont {Nagaosa}},\ }\bibfield  {title} {\bibinfo {title} {Shift current from electromagnon excitations in multiferroics},\ }\href {https://doi.org/10.1103/PhysRevB.100.235138} {\bibfield  {journal} {\bibinfo  {journal} {Physical Review B}\ }\textbf {\bibinfo {volume} {100}},\ \bibinfo {pages} {235138} (\bibinfo {year} {2019})}\BibitemShut {NoStop}%
\bibitem [{\citenamefont {Morimoto}\ \emph {et~al.}(2021)\citenamefont {Morimoto}, \citenamefont {Kitamura},\ and\ \citenamefont {Okumura}}]{Morimoto2021}%
  \BibitemOpen
  \bibfield  {author} {\bibinfo {author} {\bibfnamefont {T.}~\bibnamefont {Morimoto}}, \bibinfo {author} {\bibfnamefont {S.}~\bibnamefont {Kitamura}},\ and\ \bibinfo {author} {\bibfnamefont {S.}~\bibnamefont {Okumura}},\ }\bibfield  {title} {\bibinfo {title} {Electric polarization and nonlinear optical effects in noncentrosymmetric magnets},\ }\href {https://doi.org/10.1103/PhysRevB.104.075139} {\bibfield  {journal} {\bibinfo  {journal} {Physical Review B}\ }\textbf {\bibinfo {volume} {104}},\ \bibinfo {pages} {075139} (\bibinfo {year} {2021})}\BibitemShut {NoStop}%
\bibitem [{\citenamefont {Toshio}\ and\ \citenamefont {Kawakami}(2022)}]{Toshio2022}%
  \BibitemOpen
  \bibfield  {author} {\bibinfo {author} {\bibfnamefont {R.}~\bibnamefont {Toshio}}\ and\ \bibinfo {author} {\bibfnamefont {N.}~\bibnamefont {Kawakami}},\ }\bibfield  {title} {\bibinfo {title} {Plasmonic quantum nonlinear hall effect in noncentrosymmetric two-dimensional materials},\ }\href {https://doi.org/10.1103/PhysRevB.106.L201301} {\bibfield  {journal} {\bibinfo  {journal} {Physical Review B}\ }\textbf {\bibinfo {volume} {106}},\ \bibinfo {pages} {L201301} (\bibinfo {year} {2022})}\BibitemShut {NoStop}%
\bibitem [{\citenamefont {Morimoto}\ and\ \citenamefont {Nagaosa}(2024)}]{Morimoto2024}%
  \BibitemOpen
  \bibfield  {author} {\bibinfo {author} {\bibfnamefont {T.}~\bibnamefont {Morimoto}}\ and\ \bibinfo {author} {\bibfnamefont {N.}~\bibnamefont {Nagaosa}},\ }\bibfield  {title} {\bibinfo {title} {Direct current generation by dielectric loss in ferroelectrics},\ }\href {https://doi.org/10.1103/PhysRevB.110.045129} {\bibfield  {journal} {\bibinfo  {journal} {Physical Review B}\ }\textbf {\bibinfo {volume} {110}},\ \bibinfo {pages} {045129} (\bibinfo {year} {2024})}\BibitemShut {NoStop}%
\bibitem [{\citenamefont {Chan}\ \emph {et~al.}(2021)\citenamefont {Chan}, \citenamefont {Qiu}, \citenamefont {da~Jornada},\ and\ \citenamefont {Louie}}]{Chan2021}%
  \BibitemOpen
  \bibfield  {author} {\bibinfo {author} {\bibfnamefont {Y.-H.}\ \bibnamefont {Chan}}, \bibinfo {author} {\bibfnamefont {D.~Y.}\ \bibnamefont {Qiu}}, \bibinfo {author} {\bibfnamefont {F.~H.}\ \bibnamefont {da~Jornada}},\ and\ \bibinfo {author} {\bibfnamefont {S.~G.}\ \bibnamefont {Louie}},\ }\bibfield  {title} {\bibinfo {title} {Giant exciton-enhanced shift currents and direct current conduction with subbandgap photo excitations produced by many-electron interactions},\ }\href {https://doi.org/10.1073/pnas.1906938118} {\bibfield  {journal} {\bibinfo  {journal} {Proceedings of the National Academy of Sciences}\ }\textbf {\bibinfo {volume} {118}},\ \bibinfo {pages} {e1906938118} (\bibinfo {year} {2021})}\BibitemShut {NoStop}%
\bibitem [{\citenamefont {Kaneko}\ \emph {et~al.}(2021)\citenamefont {Kaneko}, \citenamefont {Sun}, \citenamefont {Murakami}, \citenamefont {Golež},\ and\ \citenamefont {Millis}}]{Kaneko2021}%
  \BibitemOpen
  \bibfield  {author} {\bibinfo {author} {\bibfnamefont {T.}~\bibnamefont {Kaneko}}, \bibinfo {author} {\bibfnamefont {Z.}~\bibnamefont {Sun}}, \bibinfo {author} {\bibfnamefont {Y.}~\bibnamefont {Murakami}}, \bibinfo {author} {\bibfnamefont {D.}~\bibnamefont {Golež}},\ and\ \bibinfo {author} {\bibfnamefont {A.~J.}\ \bibnamefont {Millis}},\ }\bibfield  {title} {\bibinfo {title} {Bulk photovoltaic effect driven by collective excitations in a correlated insulator},\ }\href {https://doi.org/10.1103/PhysRevLett.127.127402} {\bibfield  {journal} {\bibinfo  {journal} {Physical Review Letters}\ }\textbf {\bibinfo {volume} {127}},\ \bibinfo {pages} {127402} (\bibinfo {year} {2021})}\BibitemShut {NoStop}%
\bibitem [{\citenamefont {Iguchi}\ \emph {et~al.}(2024)\citenamefont {Iguchi}, \citenamefont {Watanabe}, \citenamefont {Murakami}, \citenamefont {Nomoto},\ and\ \citenamefont {Arita}}]{Iguchi2024}%
  \BibitemOpen
  \bibfield  {author} {\bibinfo {author} {\bibfnamefont {J.}~\bibnamefont {Iguchi}}, \bibinfo {author} {\bibfnamefont {H.}~\bibnamefont {Watanabe}}, \bibinfo {author} {\bibfnamefont {Y.}~\bibnamefont {Murakami}}, \bibinfo {author} {\bibfnamefont {T.}~\bibnamefont {Nomoto}},\ and\ \bibinfo {author} {\bibfnamefont {R.}~\bibnamefont {Arita}},\ }\bibfield  {title} {\bibinfo {title} {Bulk photovoltaic effect in antiferromagnet: Role of collective spin dynamics},\ }\href {https://doi.org/10.1103/PhysRevB.109.064407} {\bibfield  {journal} {\bibinfo  {journal} {Physical Review B}\ }\textbf {\bibinfo {volume} {109}},\ \bibinfo {pages} {064407} (\bibinfo {year} {2024})}\BibitemShut {NoStop}%
\bibitem [{\citenamefont {Hattori}\ \emph {et~al.}(2025)\citenamefont {Hattori}, \citenamefont {Watanabe},\ and\ \citenamefont {Arita}}]{hattori2025}%
  \BibitemOpen
  \bibfield  {author} {\bibinfo {author} {\bibfnamefont {K.}~\bibnamefont {Hattori}}, \bibinfo {author} {\bibfnamefont {H.}~\bibnamefont {Watanabe}},\ and\ \bibinfo {author} {\bibfnamefont {R.}~\bibnamefont {Arita}},\ }\bibfield  {title} {\bibinfo {title} {Nonlinear hall effect driven by spin-charge-coupled motive force},\ }\href {https://doi.org/10.1103/PhysRevB.111.174416} {\bibfield  {journal} {\bibinfo  {journal} {Physical Review B}\ }\textbf {\bibinfo {volume} {111}},\ \bibinfo {pages} {174416} (\bibinfo {year} {2025})}\BibitemShut {NoStop}%
\bibitem [{\citenamefont {Sotome}\ \emph {et~al.}(2019)\citenamefont {Sotome}, \citenamefont {Nakamura}, \citenamefont {Fujioka}, \citenamefont {Ogino}, \citenamefont {Kaneko}, \citenamefont {Morimoto}, \citenamefont {Zhang}, \citenamefont {Kawasaki}, \citenamefont {Nagaosa}, \citenamefont {Tokura},\ and\ \citenamefont {Ogawa}}]{Sotome2019}%
  \BibitemOpen
  \bibfield  {author} {\bibinfo {author} {\bibfnamefont {M.}~\bibnamefont {Sotome}}, \bibinfo {author} {\bibfnamefont {M.}~\bibnamefont {Nakamura}}, \bibinfo {author} {\bibfnamefont {J.}~\bibnamefont {Fujioka}}, \bibinfo {author} {\bibfnamefont {M.}~\bibnamefont {Ogino}}, \bibinfo {author} {\bibfnamefont {Y.}~\bibnamefont {Kaneko}}, \bibinfo {author} {\bibfnamefont {T.}~\bibnamefont {Morimoto}}, \bibinfo {author} {\bibfnamefont {Y.}~\bibnamefont {Zhang}}, \bibinfo {author} {\bibfnamefont {M.}~\bibnamefont {Kawasaki}}, \bibinfo {author} {\bibfnamefont {N.}~\bibnamefont {Nagaosa}}, \bibinfo {author} {\bibfnamefont {Y.}~\bibnamefont {Tokura}},\ and\ \bibinfo {author} {\bibfnamefont {N.}~\bibnamefont {Ogawa}},\ }\bibfield  {title} {\bibinfo {title} {Spectral dynamics of shift current in ferroelectric semiconductor sbsi},\ }\href {https://doi.org/10.1073/pnas.1802427116} {\bibfield  {journal} {\bibinfo  {journal} {Proceedings of the National Academy of Sciences}\ }\textbf {\bibinfo {volume} {116}},\ \bibinfo
  {pages} {1929} (\bibinfo {year} {2019})}\BibitemShut {NoStop}%
\bibitem [{\citenamefont {Sotome}\ \emph {et~al.}(2021)\citenamefont {Sotome}, \citenamefont {Nakamura}, \citenamefont {Morimoto}, \citenamefont {Zhang}, \citenamefont {Guo}, \citenamefont {Kawasaki}, \citenamefont {Nagaosa}, \citenamefont {Tokura},\ and\ \citenamefont {Ogawa}}]{Sotome2021}%
  \BibitemOpen
  \bibfield  {author} {\bibinfo {author} {\bibfnamefont {M.}~\bibnamefont {Sotome}}, \bibinfo {author} {\bibfnamefont {M.}~\bibnamefont {Nakamura}}, \bibinfo {author} {\bibfnamefont {T.}~\bibnamefont {Morimoto}}, \bibinfo {author} {\bibfnamefont {Y.}~\bibnamefont {Zhang}}, \bibinfo {author} {\bibfnamefont {G.-Y.}\ \bibnamefont {Guo}}, \bibinfo {author} {\bibfnamefont {M.}~\bibnamefont {Kawasaki}}, \bibinfo {author} {\bibfnamefont {N.}~\bibnamefont {Nagaosa}}, \bibinfo {author} {\bibfnamefont {Y.}~\bibnamefont {Tokura}},\ and\ \bibinfo {author} {\bibfnamefont {N.}~\bibnamefont {Ogawa}},\ }\bibfield  {title} {\bibinfo {title} {Terahertz emission spectroscopy of ultrafast exciton shift current in the noncentrosymmetric semiconductor cds},\ }\href {https://doi.org/10.1103/PhysRevB.103.L241111} {\bibfield  {journal} {\bibinfo  {journal} {Physical Review B}\ }\textbf {\bibinfo {volume} {103}},\ \bibinfo {pages} {L241111} (\bibinfo {year} {2021})}\BibitemShut {NoStop}%
\bibitem [{\citenamefont {Nakamura}\ \emph {et~al.}(2024)\citenamefont {Nakamura}, \citenamefont {Chan}, \citenamefont {Yasunami}, \citenamefont {Huang}, \citenamefont {Guo}, \citenamefont {Hu}, \citenamefont {Ogawa}, \citenamefont {Chiew}, \citenamefont {Yu}, \citenamefont {Morimoto}, \citenamefont {Nagaosa}, \citenamefont {Tokura},\ and\ \citenamefont {Kawasaki}}]{Nakamura2024}%
  \BibitemOpen
  \bibfield  {author} {\bibinfo {author} {\bibfnamefont {M.}~\bibnamefont {Nakamura}}, \bibinfo {author} {\bibfnamefont {Y.-H.}\ \bibnamefont {Chan}}, \bibinfo {author} {\bibfnamefont {T.}~\bibnamefont {Yasunami}}, \bibinfo {author} {\bibfnamefont {Y.-S.}\ \bibnamefont {Huang}}, \bibinfo {author} {\bibfnamefont {G.-Y.}\ \bibnamefont {Guo}}, \bibinfo {author} {\bibfnamefont {Y.}~\bibnamefont {Hu}}, \bibinfo {author} {\bibfnamefont {N.}~\bibnamefont {Ogawa}}, \bibinfo {author} {\bibfnamefont {Y.}~\bibnamefont {Chiew}}, \bibinfo {author} {\bibfnamefont {X.}~\bibnamefont {Yu}}, \bibinfo {author} {\bibfnamefont {T.}~\bibnamefont {Morimoto}}, \bibinfo {author} {\bibfnamefont {N.}~\bibnamefont {Nagaosa}}, \bibinfo {author} {\bibfnamefont {Y.}~\bibnamefont {Tokura}},\ and\ \bibinfo {author} {\bibfnamefont {M.}~\bibnamefont {Kawasaki}},\ }\bibfield  {title} {\bibinfo {title} {Strongly enhanced shift current at exciton resonances in a noncentrosymmetric wide-gap semiconductor},\ }\href
  {https://doi.org/10.1038/s41467-024-53541-6} {\bibfield  {journal} {\bibinfo  {journal} {Nature Communications}\ }\textbf {\bibinfo {volume} {15}},\ \bibinfo {pages} {9672} (\bibinfo {year} {2024})}\BibitemShut {NoStop}%
\bibitem [{\citenamefont {Okamura}\ \emph {et~al.}(2022)\citenamefont {Okamura}, \citenamefont {Morimoto}, \citenamefont {Ogawa}, \citenamefont {Kaneko}, \citenamefont {Guo}, \citenamefont {Nakamura}, \citenamefont {Kawasaki}, \citenamefont {Nagaosa}, \citenamefont {Tokura},\ and\ \citenamefont {Takahashi}}]{Okamura2022}%
  \BibitemOpen
  \bibfield  {author} {\bibinfo {author} {\bibfnamefont {Y.}~\bibnamefont {Okamura}}, \bibinfo {author} {\bibfnamefont {T.}~\bibnamefont {Morimoto}}, \bibinfo {author} {\bibfnamefont {N.}~\bibnamefont {Ogawa}}, \bibinfo {author} {\bibfnamefont {Y.}~\bibnamefont {Kaneko}}, \bibinfo {author} {\bibfnamefont {G.-Y.}\ \bibnamefont {Guo}}, \bibinfo {author} {\bibfnamefont {M.}~\bibnamefont {Nakamura}}, \bibinfo {author} {\bibfnamefont {M.}~\bibnamefont {Kawasaki}}, \bibinfo {author} {\bibfnamefont {N.}~\bibnamefont {Nagaosa}}, \bibinfo {author} {\bibfnamefont {Y.}~\bibnamefont {Tokura}},\ and\ \bibinfo {author} {\bibfnamefont {Y.}~\bibnamefont {Takahashi}},\ }\bibfield  {title} {\bibinfo {title} {Photovoltaic effect by soft phonon excitation},\ }\bibfield  {journal} {\bibinfo  {journal} {Proceedings of the National Academy of Sciences}\ }\textbf {\bibinfo {volume} {119}},\ \href {https://doi.org/10.1073/pnas.2122313119} {10.1073/pnas.2122313119} (\bibinfo {year} {2022})\BibitemShut {NoStop}%
\bibitem [{\citenamefont {Ogino}\ \emph {et~al.}(2024)\citenamefont {Ogino}, \citenamefont {Okamura}, \citenamefont {Fujiwara}, \citenamefont {Morimoto}, \citenamefont {Nagaosa}, \citenamefont {Kaneko}, \citenamefont {Tokura},\ and\ \citenamefont {Takahashi}}]{Ogino2024}%
  \BibitemOpen
  \bibfield  {author} {\bibinfo {author} {\bibfnamefont {M.}~\bibnamefont {Ogino}}, \bibinfo {author} {\bibfnamefont {Y.}~\bibnamefont {Okamura}}, \bibinfo {author} {\bibfnamefont {K.}~\bibnamefont {Fujiwara}}, \bibinfo {author} {\bibfnamefont {T.}~\bibnamefont {Morimoto}}, \bibinfo {author} {\bibfnamefont {N.}~\bibnamefont {Nagaosa}}, \bibinfo {author} {\bibfnamefont {Y.}~\bibnamefont {Kaneko}}, \bibinfo {author} {\bibfnamefont {Y.}~\bibnamefont {Tokura}},\ and\ \bibinfo {author} {\bibfnamefont {Y.}~\bibnamefont {Takahashi}},\ }\bibfield  {title} {\bibinfo {title} {Terahertz photon to dc current conversion via magnetic excitations of multiferroics},\ }\href {https://doi.org/10.1038/s41467-024-49056-9} {\bibfield  {journal} {\bibinfo  {journal} {Nature Communications}\ }\textbf {\bibinfo {volume} {15}},\ \bibinfo {pages} {4699} (\bibinfo {year} {2024})}\BibitemShut {NoStop}%
\bibitem [{\citenamefont {Jungwirth}\ \emph {et~al.}(2016)\citenamefont {Jungwirth}, \citenamefont {Marti}, \citenamefont {Wadley},\ and\ \citenamefont {Wunderlich}}]{Jungwirth_review2016}%
  \BibitemOpen
  \bibfield  {author} {\bibinfo {author} {\bibfnamefont {T.}~\bibnamefont {Jungwirth}}, \bibinfo {author} {\bibfnamefont {X.}~\bibnamefont {Marti}}, \bibinfo {author} {\bibfnamefont {P.}~\bibnamefont {Wadley}},\ and\ \bibinfo {author} {\bibfnamefont {J.}~\bibnamefont {Wunderlich}},\ }\bibfield  {title} {\bibinfo {title} {Antiferromagnetic spintronics},\ }\href {https://doi.org/10.1038/nnano.2016.18} {\bibfield  {journal} {\bibinfo  {journal} {Nature Nanotechnology}\ }\textbf {\bibinfo {volume} {11}},\ \bibinfo {pages} {231} (\bibinfo {year} {2016})}\BibitemShut {NoStop}%
\bibitem [{\citenamefont {Baltz}\ \emph {et~al.}(2018)\citenamefont {Baltz}, \citenamefont {Manchon}, \citenamefont {Tsoi}, \citenamefont {Moriyama}, \citenamefont {Ono},\ and\ \citenamefont {Tserkovnyak}}]{Baltz2018}%
  \BibitemOpen
  \bibfield  {author} {\bibinfo {author} {\bibfnamefont {V.}~\bibnamefont {Baltz}}, \bibinfo {author} {\bibfnamefont {A.}~\bibnamefont {Manchon}}, \bibinfo {author} {\bibfnamefont {M.}~\bibnamefont {Tsoi}}, \bibinfo {author} {\bibfnamefont {T.}~\bibnamefont {Moriyama}}, \bibinfo {author} {\bibfnamefont {T.}~\bibnamefont {Ono}},\ and\ \bibinfo {author} {\bibfnamefont {Y.}~\bibnamefont {Tserkovnyak}},\ }\bibfield  {title} {\bibinfo {title} {Antiferromagnetic spintronics},\ }\href {https://doi.org/10.1103/RevModPhys.90.015005} {\bibfield  {journal} {\bibinfo  {journal} {Review of Modern Physics}\ }\textbf {\bibinfo {volume} {90}},\ \bibinfo {pages} {015005} (\bibinfo {year} {2018})}\BibitemShut {NoStop}%
\bibitem [{\citenamefont {Manchon}\ \emph {et~al.}(2019)\citenamefont {Manchon}, \citenamefont {\ifmmode~\check{Z}\else \v{Z}\fi{}elezn\'y}, \citenamefont {Miron}, \citenamefont {Jungwirth}, \citenamefont {Sinova}, \citenamefont {Thiaville}, \citenamefont {Garello},\ and\ \citenamefont {Gambardella}}]{Manchon2019}%
  \BibitemOpen
  \bibfield  {author} {\bibinfo {author} {\bibfnamefont {A.}~\bibnamefont {Manchon}}, \bibinfo {author} {\bibfnamefont {J.}~\bibnamefont {\ifmmode~\check{Z}\else \v{Z}\fi{}elezn\'y}}, \bibinfo {author} {\bibfnamefont {I.~M.}\ \bibnamefont {Miron}}, \bibinfo {author} {\bibfnamefont {T.}~\bibnamefont {Jungwirth}}, \bibinfo {author} {\bibfnamefont {J.}~\bibnamefont {Sinova}}, \bibinfo {author} {\bibfnamefont {A.}~\bibnamefont {Thiaville}}, \bibinfo {author} {\bibfnamefont {K.}~\bibnamefont {Garello}},\ and\ \bibinfo {author} {\bibfnamefont {P.}~\bibnamefont {Gambardella}},\ }\bibfield  {title} {\bibinfo {title} {Current-induced spin-orbit torques in ferromagnetic and antiferromagnetic systems},\ }\href {https://doi.org/10.1103/RevModPhys.91.035004} {\bibfield  {journal} {\bibinfo  {journal} {Review of Modern Physics}\ }\textbf {\bibinfo {volume} {91}},\ \bibinfo {pages} {035004} (\bibinfo {year} {2019})}\BibitemShut {NoStop}%
\bibitem [{\citenamefont {{Levitov}}\ \emph {et~al.}(1985)\citenamefont {{Levitov}}, \citenamefont {{Nazarov}},\ and\ \citenamefont {{Eliashberg}}}]{Levitov1985}%
  \BibitemOpen
  \bibfield  {author} {\bibinfo {author} {\bibfnamefont {L.~S.}\ \bibnamefont {{Levitov}}}, \bibinfo {author} {\bibfnamefont {Y.~V.}\ \bibnamefont {{Nazarov}}},\ and\ \bibinfo {author} {\bibfnamefont {G.~M.}\ \bibnamefont {{Eliashberg}}},\ }\bibfield  {title} {\bibinfo {title} {{Magnetoelectric effects in conductors with mirror isomer symmetry}},\ }\href@noop {} {\bibfield  {journal} {\bibinfo  {journal} {Soviet Journal of Experimental and Theoretical Physics}\ }\textbf {\bibinfo {volume} {61}},\ \bibinfo {pages} {133} (\bibinfo {year} {1985})}\BibitemShut {NoStop}%
\bibitem [{\citenamefont {Edelstein}(1990)}]{Edelstein1990}%
  \BibitemOpen
  \bibfield  {author} {\bibinfo {author} {\bibfnamefont {V.}~\bibnamefont {Edelstein}},\ }\bibfield  {title} {\bibinfo {title} {Spin polarization of conduction electrons induced by electric current in two-dimensional asymmetric electron systems},\ }\href {https://doi.org/https://doi.org/10.1016/0038-1098(90)90963-C} {\bibfield  {journal} {\bibinfo  {journal} {Solid State Communications}\ }\textbf {\bibinfo {volume} {73}},\ \bibinfo {pages} {233} (\bibinfo {year} {1990})}\BibitemShut {NoStop}%
\bibitem [{\citenamefont {Yanase}(2014)}]{Yanase2014}%
  \BibitemOpen
  \bibfield  {author} {\bibinfo {author} {\bibfnamefont {Y.}~\bibnamefont {Yanase}},\ }\bibfield  {title} {\bibinfo {title} {Magneto-electric effect in three-dimensional coupled zigzag chains},\ }\href {https://doi.org/10.7566/JPSJ.83.014703} {\bibfield  {journal} {\bibinfo  {journal} {Journal of the Physical Society of Japan}\ }\textbf {\bibinfo {volume} {83}},\ \bibinfo {pages} {014703} (\bibinfo {year} {2014})}\BibitemShut {NoStop}%
\bibitem [{\citenamefont {\ifmmode~\check{Z}\else \v{Z}\fi{}elezn\'y}\ \emph {et~al.}(2014)\citenamefont {\ifmmode~\check{Z}\else \v{Z}\fi{}elezn\'y}, \citenamefont {Gao}, \citenamefont {V\'yborn\'y}, \citenamefont {Zemen}, \citenamefont {Ma\ifmmode~\check{s}\else \v{s}\fi{}ek}, \citenamefont {Manchon}, \citenamefont {Wunderlich}, \citenamefont {Sinova},\ and\ \citenamefont {Jungwirth}}]{Zelezny2014}%
  \BibitemOpen
  \bibfield  {author} {\bibinfo {author} {\bibfnamefont {J.}~\bibnamefont {\ifmmode~\check{Z}\else \v{Z}\fi{}elezn\'y}}, \bibinfo {author} {\bibfnamefont {H.}~\bibnamefont {Gao}}, \bibinfo {author} {\bibfnamefont {K.}~\bibnamefont {V\'yborn\'y}}, \bibinfo {author} {\bibfnamefont {J.}~\bibnamefont {Zemen}}, \bibinfo {author} {\bibfnamefont {J.}~\bibnamefont {Ma\ifmmode~\check{s}\else \v{s}\fi{}ek}}, \bibinfo {author} {\bibfnamefont {A.}~\bibnamefont {Manchon}}, \bibinfo {author} {\bibfnamefont {J.}~\bibnamefont {Wunderlich}}, \bibinfo {author} {\bibfnamefont {J.}~\bibnamefont {Sinova}},\ and\ \bibinfo {author} {\bibfnamefont {T.}~\bibnamefont {Jungwirth}},\ }\bibfield  {title} {\bibinfo {title} {Relativistic n\'eel-order fields induced by electrical current in antiferromagnets},\ }\href {https://doi.org/10.1103/PhysRevLett.113.157201} {\bibfield  {journal} {\bibinfo  {journal} {Physical Review Letter}\ }\textbf {\bibinfo {volume} {113}},\ \bibinfo {pages} {157201} (\bibinfo {year} {2014})}\BibitemShut
  {NoStop}%
\bibitem [{\citenamefont {Niu}\ and\ \citenamefont {Kleinman}(1998)}]{Niu1998}%
  \BibitemOpen
  \bibfield  {author} {\bibinfo {author} {\bibfnamefont {Q.}~\bibnamefont {Niu}}\ and\ \bibinfo {author} {\bibfnamefont {L.}~\bibnamefont {Kleinman}},\ }\bibfield  {title} {\bibinfo {title} {Spin-wave dynamics in real crystals},\ }\href {https://doi.org/10.1103/PhysRevLett.80.2205} {\bibfield  {journal} {\bibinfo  {journal} {Phys. Rev. Lett.}\ }\textbf {\bibinfo {volume} {80}},\ \bibinfo {pages} {2205} (\bibinfo {year} {1998})}\BibitemShut {NoStop}%
\bibitem [{\citenamefont {Niu}\ \emph {et~al.}(1999)\citenamefont {Niu}, \citenamefont {Wang}, \citenamefont {Kleinman}, \citenamefont {Liu}, \citenamefont {Nicholson},\ and\ \citenamefont {Stocks}}]{Niu1999}%
  \BibitemOpen
  \bibfield  {author} {\bibinfo {author} {\bibfnamefont {Q.}~\bibnamefont {Niu}}, \bibinfo {author} {\bibfnamefont {X.}~\bibnamefont {Wang}}, \bibinfo {author} {\bibfnamefont {L.}~\bibnamefont {Kleinman}}, \bibinfo {author} {\bibfnamefont {W.-M.}\ \bibnamefont {Liu}}, \bibinfo {author} {\bibfnamefont {D.~M.~C.}\ \bibnamefont {Nicholson}},\ and\ \bibinfo {author} {\bibfnamefont {G.~M.}\ \bibnamefont {Stocks}},\ }\bibfield  {title} {\bibinfo {title} {Adiabatic dynamics of local spin moments in itinerant magnets},\ }\href {https://doi.org/10.1103/PhysRevLett.83.207} {\bibfield  {journal} {\bibinfo  {journal} {Phys. Rev. Lett.}\ }\textbf {\bibinfo {volume} {83}},\ \bibinfo {pages} {207} (\bibinfo {year} {1999})}\BibitemShut {NoStop}%
\bibitem [{\citenamefont {Stahl}\ and\ \citenamefont {Potthoff}(2017)}]{Cristopher2017}%
  \BibitemOpen
  \bibfield  {author} {\bibinfo {author} {\bibfnamefont {C.}~\bibnamefont {Stahl}}\ and\ \bibinfo {author} {\bibfnamefont {M.}~\bibnamefont {Potthoff}},\ }\bibfield  {title} {\bibinfo {title} {Anomalous spin precession under a geometrical torque},\ }\href {https://doi.org/10.1103/PhysRevLett.119.227203} {\bibfield  {journal} {\bibinfo  {journal} {Phys. Rev. Lett.}\ }\textbf {\bibinfo {volume} {119}},\ \bibinfo {pages} {227203} (\bibinfo {year} {2017})}\BibitemShut {NoStop}%
\bibitem [{\citenamefont {Michel}\ and\ \citenamefont {Potthoff}(2022)}]{Simon2022}%
  \BibitemOpen
  \bibfield  {author} {\bibinfo {author} {\bibfnamefont {S.}~\bibnamefont {Michel}}\ and\ \bibinfo {author} {\bibfnamefont {M.}~\bibnamefont {Potthoff}},\ }\bibfield  {title} {\bibinfo {title} {Spin berry curvature of the haldane model},\ }\href {https://doi.org/10.1103/PhysRevB.106.235423} {\bibfield  {journal} {\bibinfo  {journal} {Phys. Rev. B}\ }\textbf {\bibinfo {volume} {106}},\ \bibinfo {pages} {235423} (\bibinfo {year} {2022})}\BibitemShut {NoStop}%
\bibitem [{\citenamefont {Lenzing}\ \emph {et~al.}(2023)\citenamefont {Lenzing}, \citenamefont {Kr\"uger},\ and\ \citenamefont {Potthoff}}]{Lenzing2023}%
  \BibitemOpen
  \bibfield  {author} {\bibinfo {author} {\bibfnamefont {N.}~\bibnamefont {Lenzing}}, \bibinfo {author} {\bibfnamefont {D.}~\bibnamefont {Kr\"uger}},\ and\ \bibinfo {author} {\bibfnamefont {M.}~\bibnamefont {Potthoff}},\ }\bibfield  {title} {\bibinfo {title} {Geometrical torque on magnetic moments coupled to a correlated antiferromagnet},\ }\href {https://doi.org/10.1103/PhysRevResearch.5.L032012} {\bibfield  {journal} {\bibinfo  {journal} {Phys. Rev. Res.}\ }\textbf {\bibinfo {volume} {5}},\ \bibinfo {pages} {L032012} (\bibinfo {year} {2023})}\BibitemShut {NoStop}%
\bibitem [{\citenamefont {Sundaram}\ and\ \citenamefont {Niu}(1999)}]{Sundaram1999}%
  \BibitemOpen
  \bibfield  {author} {\bibinfo {author} {\bibfnamefont {G.}~\bibnamefont {Sundaram}}\ and\ \bibinfo {author} {\bibfnamefont {Q.}~\bibnamefont {Niu}},\ }\bibfield  {title} {\bibinfo {title} {Wave-packet dynamics in slowly perturbed crystals: Gradient corrections and berry-phase effects},\ }\href {https://doi.org/10.1103/PhysRevB.59.14915} {\bibfield  {journal} {\bibinfo  {journal} {Physical Review B}\ }\textbf {\bibinfo {volume} {59}},\ \bibinfo {pages} {14915} (\bibinfo {year} {1999})}\BibitemShut {NoStop}%
\bibitem [{\citenamefont {Xiao}\ \emph {et~al.}(2005)\citenamefont {Xiao}, \citenamefont {Shi},\ and\ \citenamefont {Niu}}]{Xiao2005}%
  \BibitemOpen
  \bibfield  {author} {\bibinfo {author} {\bibfnamefont {D.}~\bibnamefont {Xiao}}, \bibinfo {author} {\bibfnamefont {J.}~\bibnamefont {Shi}},\ and\ \bibinfo {author} {\bibfnamefont {Q.}~\bibnamefont {Niu}},\ }\bibfield  {title} {\bibinfo {title} {Berry phase correction to electron density of states in solids},\ }\href {https://doi.org/10.1103/PhysRevLett.95.137204} {\bibfield  {journal} {\bibinfo  {journal} {Physical Review Letter}\ }\textbf {\bibinfo {volume} {95}},\ \bibinfo {pages} {137204} (\bibinfo {year} {2005})}\BibitemShut {NoStop}%
\bibitem [{\citenamefont {Freimuth}\ \emph {et~al.}(2013)\citenamefont {Freimuth}, \citenamefont {Bamler}, \citenamefont {Mokrousov},\ and\ \citenamefont {Rosch}}]{Freimuth2013}%
  \BibitemOpen
  \bibfield  {author} {\bibinfo {author} {\bibfnamefont {F.}~\bibnamefont {Freimuth}}, \bibinfo {author} {\bibfnamefont {R.}~\bibnamefont {Bamler}}, \bibinfo {author} {\bibfnamefont {Y.}~\bibnamefont {Mokrousov}},\ and\ \bibinfo {author} {\bibfnamefont {A.}~\bibnamefont {Rosch}},\ }\bibfield  {title} {\bibinfo {title} {Phase-space berry phases in chiral magnets: Dzyaloshinskii-moriya interaction and the charge of skyrmions},\ }\href {https://doi.org/10.1103/PhysRevB.88.214409} {\bibfield  {journal} {\bibinfo  {journal} {Physical Review B}\ }\textbf {\bibinfo {volume} {88}},\ \bibinfo {pages} {214409} (\bibinfo {year} {2013})}\BibitemShut {NoStop}%
\bibitem [{\citenamefont {Freimuth}\ \emph {et~al.}(2014)\citenamefont {Freimuth}, \citenamefont {Bl\"ugel},\ and\ \citenamefont {Mokrousov}}]{Freimuth2014}%
  \BibitemOpen
  \bibfield  {author} {\bibinfo {author} {\bibfnamefont {F.}~\bibnamefont {Freimuth}}, \bibinfo {author} {\bibfnamefont {S.}~\bibnamefont {Bl\"ugel}},\ and\ \bibinfo {author} {\bibfnamefont {Y.}~\bibnamefont {Mokrousov}},\ }\bibfield  {title} {\bibinfo {title} {Spin-orbit torques in co/pt(111) and mn/w(001) magnetic bilayers from first principles},\ }\href {https://doi.org/10.1103/PhysRevB.90.174423} {\bibfield  {journal} {\bibinfo  {journal} {Physical Review B}\ }\textbf {\bibinfo {volume} {90}},\ \bibinfo {pages} {174423} (\bibinfo {year} {2014})}\BibitemShut {NoStop}%
\bibitem [{\citenamefont {Hanke}\ \emph {et~al.}(2017)\citenamefont {Hanke}, \citenamefont {Freimuth}, \citenamefont {Niu}, \citenamefont {Bl{\"u}gel},\ and\ \citenamefont {Mokrousov}}]{Hanke2017}%
  \BibitemOpen
  \bibfield  {author} {\bibinfo {author} {\bibfnamefont {J.-P.}\ \bibnamefont {Hanke}}, \bibinfo {author} {\bibfnamefont {F.}~\bibnamefont {Freimuth}}, \bibinfo {author} {\bibfnamefont {C.}~\bibnamefont {Niu}}, \bibinfo {author} {\bibfnamefont {S.}~\bibnamefont {Bl{\"u}gel}},\ and\ \bibinfo {author} {\bibfnamefont {Y.}~\bibnamefont {Mokrousov}},\ }\bibfield  {title} {\bibinfo {title} {Mixed weyl semimetals and low-dissipation magnetization control in insulators by spin--orbit torques},\ }\href {https://doi.org/10.1038/s41467-017-01138-7} {\bibfield  {journal} {\bibinfo  {journal} {Nature Communications}\ }\textbf {\bibinfo {volume} {8}},\ \bibinfo {pages} {1479} (\bibinfo {year} {2017})}\BibitemShut {NoStop}%
\bibitem [{\citenamefont {Xiao}\ \emph {et~al.}(2021)\citenamefont {Xiao}, \citenamefont {Xiong},\ and\ \citenamefont {Niu}}]{Xiao2021}%
  \BibitemOpen
  \bibfield  {author} {\bibinfo {author} {\bibfnamefont {C.}~\bibnamefont {Xiao}}, \bibinfo {author} {\bibfnamefont {B.}~\bibnamefont {Xiong}},\ and\ \bibinfo {author} {\bibfnamefont {Q.}~\bibnamefont {Niu}},\ }\bibfield  {title} {\bibinfo {title} {Electric driving of magnetization dynamics in a hybrid insulator},\ }\href {https://doi.org/10.1103/PhysRevB.104.064433} {\bibfield  {journal} {\bibinfo  {journal} {Phys. Rev. B}\ }\textbf {\bibinfo {volume} {104}},\ \bibinfo {pages} {064433} (\bibinfo {year} {2021})}\BibitemShut {NoStop}%
\bibitem [{\citenamefont {Meguro}\ \emph {et~al.}(2025)\citenamefont {Meguro}, \citenamefont {Ozawa}, \citenamefont {Kobayashi}, \citenamefont {Araki},\ and\ \citenamefont {Nomura}}]{Meguro2025}%
  \BibitemOpen
  \bibfield  {author} {\bibinfo {author} {\bibfnamefont {T.}~\bibnamefont {Meguro}}, \bibinfo {author} {\bibfnamefont {A.}~\bibnamefont {Ozawa}}, \bibinfo {author} {\bibfnamefont {K.}~\bibnamefont {Kobayashi}}, \bibinfo {author} {\bibfnamefont {Y.}~\bibnamefont {Araki}},\ and\ \bibinfo {author} {\bibfnamefont {K.}~\bibnamefont {Nomura}},\ }\bibfield  {title} {\bibinfo {title} {Topological spin-orbit torque in ferrimagnetic weyl semimetal},\ }\href {https://doi.org/10.1103/PhysRevResearch.7.L022065} {\bibfield  {journal} {\bibinfo  {journal} {Physical Review Research}\ }\textbf {\bibinfo {volume} {7}},\ \bibinfo {pages} {L022065} (\bibinfo {year} {2025})}\BibitemShut {NoStop}%
\bibitem [{\citenamefont {Simon}(1983)}]{Simon1983}%
  \BibitemOpen
  \bibfield  {author} {\bibinfo {author} {\bibfnamefont {B.}~\bibnamefont {Simon}},\ }\bibfield  {title} {\bibinfo {title} {Holonomy, the quantum adiabatic theorem, and berry's phase},\ }\href {https://doi.org/10.1103/PhysRevLett.51.2167} {\bibfield  {journal} {\bibinfo  {journal} {Physical Review Letters}\ }\textbf {\bibinfo {volume} {51}},\ \bibinfo {pages} {2167} (\bibinfo {year} {1983})}\BibitemShut {NoStop}%
\bibitem [{\citenamefont {Berry}(1984)}]{Berry1984}%
  \BibitemOpen
  \bibfield  {author} {\bibinfo {author} {\bibfnamefont {M.~V.}\ \bibnamefont {Berry}},\ }\bibfield  {title} {\bibinfo {title} {Quantal phase factors accompanying adiabatic changes},\ }\href {http://www.jstor.org/stable/2397741} {\bibfield  {journal} {\bibinfo  {journal} {Proceedings of the Royal Society of London. Series A, Mathematical and Physical Sciences}\ }\textbf {\bibinfo {volume} {392}},\ \bibinfo {pages} {45} (\bibinfo {year} {1984})}\BibitemShut {NoStop}%
\bibitem [{\citenamefont {Resta}(2011)}]{Resta2011}%
  \BibitemOpen
  \bibfield  {author} {\bibinfo {author} {\bibfnamefont {R.}~\bibnamefont {Resta}},\ }\bibfield  {title} {\bibinfo {title} {The insulating state of matter: a geometrical theory},\ }\href {https://doi.org/10.1140/epjb/e2010-10874-4} {\bibfield  {journal} {\bibinfo  {journal} {The European Physical Journal B}\ }\textbf {\bibinfo {volume} {79}},\ \bibinfo {pages} {121} (\bibinfo {year} {2011})}\BibitemShut {NoStop}%
\bibitem [{\citenamefont {Ahn}\ \emph {et~al.}(2022)\citenamefont {Ahn}, \citenamefont {Guo}, \citenamefont {Nagaosa},\ and\ \citenamefont {Vishwanath}}]{Ahn2022}%
  \BibitemOpen
  \bibfield  {author} {\bibinfo {author} {\bibfnamefont {J.}~\bibnamefont {Ahn}}, \bibinfo {author} {\bibfnamefont {G.-Y.}\ \bibnamefont {Guo}}, \bibinfo {author} {\bibfnamefont {N.}~\bibnamefont {Nagaosa}},\ and\ \bibinfo {author} {\bibfnamefont {A.}~\bibnamefont {Vishwanath}},\ }\bibfield  {title} {\bibinfo {title} {Riemannian geometry of resonant optical responses},\ }\href {https://doi.org/10.1038/s41567-021-01465-z} {\bibfield  {journal} {\bibinfo  {journal} {Nature Physics}\ }\textbf {\bibinfo {volume} {18}},\ \bibinfo {pages} {290} (\bibinfo {year} {2022})}\BibitemShut {NoStop}%
\bibitem [{\citenamefont {Yu}\ \emph {et~al.}(2025)\citenamefont {Yu}, \citenamefont {Bernevig}, \citenamefont {Queiroz}, \citenamefont {Rossi}, \citenamefont {Törmä},\ and\ \citenamefont {Yang}}]{Yu2025}%
  \BibitemOpen
  \bibfield  {author} {\bibinfo {author} {\bibfnamefont {J.}~\bibnamefont {Yu}}, \bibinfo {author} {\bibfnamefont {B.~A.}\ \bibnamefont {Bernevig}}, \bibinfo {author} {\bibfnamefont {R.}~\bibnamefont {Queiroz}}, \bibinfo {author} {\bibfnamefont {E.}~\bibnamefont {Rossi}}, \bibinfo {author} {\bibfnamefont {P.}~\bibnamefont {Törmä}},\ and\ \bibinfo {author} {\bibfnamefont {B.-J.}\ \bibnamefont {Yang}},\ }\href {https://arxiv.org/abs/2501.00098} {\bibinfo {title} {Quantum geometry in quantum materials}} (\bibinfo {year} {2025}),\ \Eprint {https://arxiv.org/abs/2501.00098} {arXiv:2501.00098 [cond-mat.mes-hall]} \BibitemShut {NoStop}%
\bibitem [{\citenamefont {Jiang}\ \emph {et~al.}(2025)\citenamefont {Jiang}, \citenamefont {Holder},\ and\ \citenamefont {Yan}}]{Jiang2025}%
  \BibitemOpen
  \bibfield  {author} {\bibinfo {author} {\bibfnamefont {Y.}~\bibnamefont {Jiang}}, \bibinfo {author} {\bibfnamefont {T.}~\bibnamefont {Holder}},\ and\ \bibinfo {author} {\bibfnamefont {B.}~\bibnamefont {Yan}},\ }\href {https://arxiv.org/abs/2503.04943} {\bibinfo {title} {Revealing quantum geometry in nonlinear quantum materials}} (\bibinfo {year} {2025}),\ \Eprint {https://arxiv.org/abs/2503.04943} {arXiv:2503.04943 [cond-mat.mes-hall]} \BibitemShut {NoStop}%
\bibitem [{\citenamefont {Avdoshkin}\ \emph {et~al.}(2025)\citenamefont {Avdoshkin}, \citenamefont {Mitscherling},\ and\ \citenamefont {Moore}}]{Avdoshkin2025}%
  \BibitemOpen
  \bibfield  {author} {\bibinfo {author} {\bibfnamefont {A.}~\bibnamefont {Avdoshkin}}, \bibinfo {author} {\bibfnamefont {J.}~\bibnamefont {Mitscherling}},\ and\ \bibinfo {author} {\bibfnamefont {J.~E.}\ \bibnamefont {Moore}},\ }\href {https://arxiv.org/abs/2409.16358} {\bibinfo {title} {The multi-state geometry of shift current and polarization}} (\bibinfo {year} {2025}),\ \Eprint {https://arxiv.org/abs/2409.16358} {arXiv:2409.16358 [cond-mat.str-el]} \BibitemShut {NoStop}%
\bibitem [{\citenamefont {Mitscherling}\ \emph {et~al.}(2025)\citenamefont {Mitscherling}, \citenamefont {Avdoshkin},\ and\ \citenamefont {Moore}}]{Mitscherling2025}%
  \BibitemOpen
  \bibfield  {author} {\bibinfo {author} {\bibfnamefont {J.}~\bibnamefont {Mitscherling}}, \bibinfo {author} {\bibfnamefont {A.}~\bibnamefont {Avdoshkin}},\ and\ \bibinfo {author} {\bibfnamefont {J.~E.}\ \bibnamefont {Moore}},\ }\href {https://arxiv.org/abs/2412.03637} {\bibinfo {title} {Gauge-invariant projector calculus for quantum state geometry and applications to observables in crystals}} (\bibinfo {year} {2025}),\ \Eprint {https://arxiv.org/abs/2412.03637} {arXiv:2412.03637 [cond-mat.str-el]} \BibitemShut {NoStop}%
\bibitem [{\citenamefont {\ifmmode~\check{S}\else \v{S}\fi{}mejkal}\ \emph {et~al.}(2017)\citenamefont {\ifmmode~\check{S}\else \v{S}\fi{}mejkal}, \citenamefont {\ifmmode~\check{Z}\else \v{Z}\fi{}elezn\'y}, \citenamefont {Sinova},\ and\ \citenamefont {Jungwirth}}]{Smejkal2017}%
  \BibitemOpen
  \bibfield  {author} {\bibinfo {author} {\bibfnamefont {L.}~\bibnamefont {\ifmmode~\check{S}\else \v{S}\fi{}mejkal}}, \bibinfo {author} {\bibfnamefont {J.}~\bibnamefont {\ifmmode~\check{Z}\else \v{Z}\fi{}elezn\'y}}, \bibinfo {author} {\bibfnamefont {J.}~\bibnamefont {Sinova}},\ and\ \bibinfo {author} {\bibfnamefont {T.}~\bibnamefont {Jungwirth}},\ }\bibfield  {title} {\bibinfo {title} {Electric control of dirac quasiparticles by spin-orbit torque in an antiferromagnet},\ }\href {https://doi.org/10.1103/PhysRevLett.118.106402} {\bibfield  {journal} {\bibinfo  {journal} {Physical Review Letters}\ }\textbf {\bibinfo {volume} {118}},\ \bibinfo {pages} {106402} (\bibinfo {year} {2017})}\BibitemShut {NoStop}%
\bibitem [{\citenamefont {Wadley}\ \emph {et~al.}(2016)\citenamefont {Wadley}, \citenamefont {Howells}, \citenamefont {Železný}, \citenamefont {Andrews}, \citenamefont {Hills}, \citenamefont {Campion}, \citenamefont {Novák}, \citenamefont {Olejník}, \citenamefont {Maccherozzi}, \citenamefont {Dhesi}, \citenamefont {Martin}, \citenamefont {Wagner}, \citenamefont {Wunderlich}, \citenamefont {Freimuth}, \citenamefont {Mokrousov}, \citenamefont {Kuneš}, \citenamefont {Chauhan}, \citenamefont {Grzybowski}, \citenamefont {Rushforth}, \citenamefont {Edmonds}, \citenamefont {Gallagher},\ and\ \citenamefont {Jungwirth}}]{Wadley2016}%
  \BibitemOpen
  \bibfield  {author} {\bibinfo {author} {\bibfnamefont {P.}~\bibnamefont {Wadley}}, \bibinfo {author} {\bibfnamefont {B.}~\bibnamefont {Howells}}, \bibinfo {author} {\bibfnamefont {J.}~\bibnamefont {Železný}}, \bibinfo {author} {\bibfnamefont {C.}~\bibnamefont {Andrews}}, \bibinfo {author} {\bibfnamefont {V.}~\bibnamefont {Hills}}, \bibinfo {author} {\bibfnamefont {R.~P.}\ \bibnamefont {Campion}}, \bibinfo {author} {\bibfnamefont {V.}~\bibnamefont {Novák}}, \bibinfo {author} {\bibfnamefont {K.}~\bibnamefont {Olejník}}, \bibinfo {author} {\bibfnamefont {F.}~\bibnamefont {Maccherozzi}}, \bibinfo {author} {\bibfnamefont {S.~S.}\ \bibnamefont {Dhesi}}, \bibinfo {author} {\bibfnamefont {S.~Y.}\ \bibnamefont {Martin}}, \bibinfo {author} {\bibfnamefont {T.}~\bibnamefont {Wagner}}, \bibinfo {author} {\bibfnamefont {J.}~\bibnamefont {Wunderlich}}, \bibinfo {author} {\bibfnamefont {F.}~\bibnamefont {Freimuth}}, \bibinfo {author} {\bibfnamefont {Y.}~\bibnamefont {Mokrousov}}, \bibinfo {author} {\bibfnamefont
  {J.}~\bibnamefont {Kuneš}}, \bibinfo {author} {\bibfnamefont {J.~S.}\ \bibnamefont {Chauhan}}, \bibinfo {author} {\bibfnamefont {M.~J.}\ \bibnamefont {Grzybowski}}, \bibinfo {author} {\bibfnamefont {A.~W.}\ \bibnamefont {Rushforth}}, \bibinfo {author} {\bibfnamefont {K.~W.}\ \bibnamefont {Edmonds}}, \bibinfo {author} {\bibfnamefont {B.~L.}\ \bibnamefont {Gallagher}},\ and\ \bibinfo {author} {\bibfnamefont {T.}~\bibnamefont {Jungwirth}},\ }\bibfield  {title} {\bibinfo {title} {Electrical switching of an antiferromagnet},\ }\href {https://doi.org/10.1126/science.aab1031} {\bibfield  {journal} {\bibinfo  {journal} {Science}\ }\textbf {\bibinfo {volume} {351}},\ \bibinfo {pages} {587} (\bibinfo {year} {2016})}\BibitemShut {NoStop}%
\bibitem [{\citenamefont {Tang}\ \emph {et~al.}(2016)\citenamefont {Tang}, \citenamefont {Zhou}, \citenamefont {Xu},\ and\ \citenamefont {Zhang}}]{Tang2016}%
  \BibitemOpen
  \bibfield  {author} {\bibinfo {author} {\bibfnamefont {P.}~\bibnamefont {Tang}}, \bibinfo {author} {\bibfnamefont {Q.}~\bibnamefont {Zhou}}, \bibinfo {author} {\bibfnamefont {G.}~\bibnamefont {Xu}},\ and\ \bibinfo {author} {\bibfnamefont {S.-C.}\ \bibnamefont {Zhang}},\ }\bibfield  {title} {\bibinfo {title} {Dirac fermions in an antiferromagnetic semimetal},\ }\href {https://doi.org/10.1038/nphys3839} {\bibfield  {journal} {\bibinfo  {journal} {Nature Physics}\ }\textbf {\bibinfo {volume} {12}},\ \bibinfo {pages} {1100} (\bibinfo {year} {2016})}\BibitemShut {NoStop}%
\bibitem [{\citenamefont {Godinho}\ \emph {et~al.}(2018)\citenamefont {Godinho}, \citenamefont {Reichlov{\'a}}, \citenamefont {Kriegner}, \citenamefont {Nov{\'a}k}, \citenamefont {Olejn{\'i}k}, \citenamefont {Ka{\v{s}}par}, \citenamefont {{\v{S}}ob{\'a}{\v{n}}}, \citenamefont {Wadley}, \citenamefont {Campion}, \citenamefont {Otxoa}, \citenamefont {Roy}, \citenamefont {{\v{Z}}elezn{\'y}}, \citenamefont {Jungwirth},\ and\ \citenamefont {Wunderlich}}]{Godinho2018}%
  \BibitemOpen
  \bibfield  {author} {\bibinfo {author} {\bibfnamefont {J.}~\bibnamefont {Godinho}}, \bibinfo {author} {\bibfnamefont {H.}~\bibnamefont {Reichlov{\'a}}}, \bibinfo {author} {\bibfnamefont {D.}~\bibnamefont {Kriegner}}, \bibinfo {author} {\bibfnamefont {V.}~\bibnamefont {Nov{\'a}k}}, \bibinfo {author} {\bibfnamefont {K.}~\bibnamefont {Olejn{\'i}k}}, \bibinfo {author} {\bibfnamefont {Z.}~\bibnamefont {Ka{\v{s}}par}}, \bibinfo {author} {\bibfnamefont {Z.}~\bibnamefont {{\v{S}}ob{\'a}{\v{n}}}}, \bibinfo {author} {\bibfnamefont {P.}~\bibnamefont {Wadley}}, \bibinfo {author} {\bibfnamefont {R.~P.}\ \bibnamefont {Campion}}, \bibinfo {author} {\bibfnamefont {R.~M.}\ \bibnamefont {Otxoa}}, \bibinfo {author} {\bibfnamefont {P.~E.}\ \bibnamefont {Roy}}, \bibinfo {author} {\bibfnamefont {J.}~\bibnamefont {{\v{Z}}elezn{\'y}}}, \bibinfo {author} {\bibfnamefont {T.}~\bibnamefont {Jungwirth}},\ and\ \bibinfo {author} {\bibfnamefont {J.}~\bibnamefont {Wunderlich}},\ }\bibfield  {title} {\bibinfo {title} {Electrically induced
  and detected n{\'e}el vector reversal in a collinear antiferromagnet},\ }\href {https://doi.org/10.1038/s41467-018-07092-2} {\bibfield  {journal} {\bibinfo  {journal} {Nature Communications}\ }\textbf {\bibinfo {volume} {9}},\ \bibinfo {pages} {4686} (\bibinfo {year} {2018})}\BibitemShut {NoStop}%
\bibitem [{\citenamefont {Linn}\ \emph {et~al.}(2023)\citenamefont {Linn}, \citenamefont {Hao}, \citenamefont {Gordon}, \citenamefont {Narayan}, \citenamefont {Berggren}, \citenamefont {Speiser}, \citenamefont {Reimers}, \citenamefont {Campion}, \citenamefont {Nov{\'a}k}, \citenamefont {Dhesi}, \citenamefont {Kim}, \citenamefont {Cacho}, \citenamefont {{\v{S}}mejkal}, \citenamefont {Jungwirth}, \citenamefont {Denlinger}, \citenamefont {Wadley},\ and\ \citenamefont {Dessau}}]{Linn2023}%
  \BibitemOpen
  \bibfield  {author} {\bibinfo {author} {\bibfnamefont {A.~G.}\ \bibnamefont {Linn}}, \bibinfo {author} {\bibfnamefont {P.}~\bibnamefont {Hao}}, \bibinfo {author} {\bibfnamefont {K.~N.}\ \bibnamefont {Gordon}}, \bibinfo {author} {\bibfnamefont {D.}~\bibnamefont {Narayan}}, \bibinfo {author} {\bibfnamefont {B.~S.}\ \bibnamefont {Berggren}}, \bibinfo {author} {\bibfnamefont {N.}~\bibnamefont {Speiser}}, \bibinfo {author} {\bibfnamefont {S.}~\bibnamefont {Reimers}}, \bibinfo {author} {\bibfnamefont {R.~P.}\ \bibnamefont {Campion}}, \bibinfo {author} {\bibfnamefont {V.}~\bibnamefont {Nov{\'a}k}}, \bibinfo {author} {\bibfnamefont {S.~S.}\ \bibnamefont {Dhesi}}, \bibinfo {author} {\bibfnamefont {T.~K.}\ \bibnamefont {Kim}}, \bibinfo {author} {\bibfnamefont {C.}~\bibnamefont {Cacho}}, \bibinfo {author} {\bibfnamefont {L.}~\bibnamefont {{\v{S}}mejkal}}, \bibinfo {author} {\bibfnamefont {T.}~\bibnamefont {Jungwirth}}, \bibinfo {author} {\bibfnamefont {J.~D.}\ \bibnamefont {Denlinger}}, \bibinfo {author}
  {\bibfnamefont {P.}~\bibnamefont {Wadley}},\ and\ \bibinfo {author} {\bibfnamefont {D.~S.}\ \bibnamefont {Dessau}},\ }\bibfield  {title} {\bibinfo {title} {Experimental electronic structure of the electrically switchable antiferromagnet cumnas},\ }\href {https://doi.org/10.1038/s41535-023-00554-x} {\bibfield  {journal} {\bibinfo  {journal} {npj Quantum Materials}\ }\textbf {\bibinfo {volume} {8}},\ \bibinfo {pages} {19} (\bibinfo {year} {2023})}\BibitemShut {NoStop}%
\bibitem [{\citenamefont {Yang}\ \emph {et~al.}(2017)\citenamefont {Yang}, \citenamefont {Bojesen}, \citenamefont {Morimoto},\ and\ \citenamefont {Furusaki}}]{Yang2017}%
  \BibitemOpen
  \bibfield  {author} {\bibinfo {author} {\bibfnamefont {B.-J.}\ \bibnamefont {Yang}}, \bibinfo {author} {\bibfnamefont {T.~A.}\ \bibnamefont {Bojesen}}, \bibinfo {author} {\bibfnamefont {T.}~\bibnamefont {Morimoto}},\ and\ \bibinfo {author} {\bibfnamefont {A.}~\bibnamefont {Furusaki}},\ }\bibfield  {title} {\bibinfo {title} {Topological semimetals protected by off-centered symmetries in nonsymmorphic crystals},\ }\href {https://doi.org/10.1103/PhysRevB.95.075135} {\bibfield  {journal} {\bibinfo  {journal} {Physical Review B}\ }\textbf {\bibinfo {volume} {95}},\ \bibinfo {pages} {075135} (\bibinfo {year} {2017})}\BibitemShut {NoStop}%
\bibitem [{sup()}]{supple}%
  \BibitemOpen
  \href@noop {} {}\bibinfo {note} {See Supplemental Material}\BibitemShut {NoStop}%
\bibitem [{\citenamefont {Ono}\ and\ \citenamefont {Ishihara}(2021)}]{Ono2021}%
  \BibitemOpen
  \bibfield  {author} {\bibinfo {author} {\bibfnamefont {A.}~\bibnamefont {Ono}}\ and\ \bibinfo {author} {\bibfnamefont {S.}~\bibnamefont {Ishihara}},\ }\bibfield  {title} {\bibinfo {title} {Ultrafast reorientation of the n{\'e}el vector in antiferromagnetic dirac semimetals},\ }\href {https://doi.org/10.1038/s41524-021-00641-2} {\bibfield  {journal} {\bibinfo  {journal} {npj Computational Materials}\ }\textbf {\bibinfo {volume} {7}},\ \bibinfo {pages} {171} (\bibinfo {year} {2021})}\BibitemShut {NoStop}%
\bibitem [{\citenamefont {Ono}\ and\ \citenamefont {Akagi}(2023)}]{ono2023}%
  \BibitemOpen
  \bibfield  {author} {\bibinfo {author} {\bibfnamefont {A.}~\bibnamefont {Ono}}\ and\ \bibinfo {author} {\bibfnamefont {Y.}~\bibnamefont {Akagi}},\ }\bibfield  {title} {\bibinfo {title} {Photocontrol of spin scalar chirality in centrosymmetric itinerant magnets},\ }\href {https://doi.org/10.1103/PhysRevB.108.L100407} {\bibfield  {journal} {\bibinfo  {journal} {Physical Review B}\ }\textbf {\bibinfo {volume} {108}},\ \bibinfo {pages} {L100407} (\bibinfo {year} {2023})}\BibitemShut {NoStop}%
\bibitem [{\citenamefont {Hattori}\ \emph {et~al.}(2024)\citenamefont {Hattori}, \citenamefont {Watanabe}, \citenamefont {Iguchi}, \citenamefont {Nomoto},\ and\ \citenamefont {Arita}}]{hattori2024}%
  \BibitemOpen
  \bibfield  {author} {\bibinfo {author} {\bibfnamefont {K.}~\bibnamefont {Hattori}}, \bibinfo {author} {\bibfnamefont {H.}~\bibnamefont {Watanabe}}, \bibinfo {author} {\bibfnamefont {J.}~\bibnamefont {Iguchi}}, \bibinfo {author} {\bibfnamefont {T.}~\bibnamefont {Nomoto}},\ and\ \bibinfo {author} {\bibfnamefont {R.}~\bibnamefont {Arita}},\ }\bibfield  {title} {\bibinfo {title} {Effect of collective spin excitations on electronic transport in topological spin textures},\ }\href {https://doi.org/10.1103/PhysRevB.110.014425} {\bibfield  {journal} {\bibinfo  {journal} {Physical Review B}\ }\textbf {\bibinfo {volume} {110}},\ \bibinfo {pages} {014425} (\bibinfo {year} {2024})}\BibitemShut {NoStop}%
\bibitem [{\citenamefont {Yue}\ and\ \citenamefont {Gaarde}(2022)}]{Yue2022}%
  \BibitemOpen
  \bibfield  {author} {\bibinfo {author} {\bibfnamefont {L.}~\bibnamefont {Yue}}\ and\ \bibinfo {author} {\bibfnamefont {M.~B.}\ \bibnamefont {Gaarde}},\ }\bibfield  {title} {\bibinfo {title} {Introduction to theory of high-harmonic generation in solids: tutorial},\ }\href {https://doi.org/10.1364/JOSAB.448602} {\bibfield  {journal} {\bibinfo  {journal} {Journal of the Optical Society of America B}\ }\textbf {\bibinfo {volume} {39}},\ \bibinfo {pages} {535} (\bibinfo {year} {2022})}\BibitemShut {NoStop}%
\bibitem [{\citenamefont {Murakami}\ and\ \citenamefont {Schüler}(2022)}]{Murakami2022}%
  \BibitemOpen
  \bibfield  {author} {\bibinfo {author} {\bibfnamefont {Y.}~\bibnamefont {Murakami}}\ and\ \bibinfo {author} {\bibfnamefont {M.}~\bibnamefont {Schüler}},\ }\bibfield  {title} {\bibinfo {title} {Doping and gap size dependence of high-harmonic generation in graphene: Importance of consistent formulation of light-matter coupling},\ }\href {https://doi.org/10.1103/PhysRevB.106.035204} {\bibfield  {journal} {\bibinfo  {journal} {Physical Review B}\ }\textbf {\bibinfo {volume} {106}},\ \bibinfo {pages} {035204} (\bibinfo {year} {2022})}\BibitemShut {NoStop}%
\bibitem [{\citenamefont {Bharti}\ and\ \citenamefont {Dixit}(2023)}]{Bharti2023}%
  \BibitemOpen
  \bibfield  {author} {\bibinfo {author} {\bibfnamefont {A.}~\bibnamefont {Bharti}}\ and\ \bibinfo {author} {\bibfnamefont {G.}~\bibnamefont {Dixit}},\ }\bibfield  {title} {\bibinfo {title} {Tailoring photocurrent in weyl semimetals via intense laser irradiation},\ }\href {https://doi.org/10.1103/PhysRevB.108.L161113} {\bibfield  {journal} {\bibinfo  {journal} {Physical Review B}\ }\textbf {\bibinfo {volume} {108}},\ \bibinfo {pages} {L161113} (\bibinfo {year} {2023})}\BibitemShut {NoStop}%
\bibitem [{\citenamefont {Bharti}\ and\ \citenamefont {Dixit}(2024)}]{Bharti2024}%
  \BibitemOpen
  \bibfield  {author} {\bibinfo {author} {\bibfnamefont {A.}~\bibnamefont {Bharti}}\ and\ \bibinfo {author} {\bibfnamefont {G.}~\bibnamefont {Dixit}},\ }\bibfield  {title} {\bibinfo {title} {Photocurrent generation in solids via linearly polarized laser},\ }\href {https://doi.org/10.1103/PhysRevB.109.104309} {\bibfield  {journal} {\bibinfo  {journal} {Physical Review B}\ }\textbf {\bibinfo {volume} {109}},\ \bibinfo {pages} {104309} (\bibinfo {year} {2024})}\BibitemShut {NoStop}%
\bibitem [{\citenamefont {Zhang}\ \emph {et~al.}(2019)\citenamefont {Zhang}, \citenamefont {Holder}, \citenamefont {Ishizuka}, \citenamefont {de~Juan}, \citenamefont {Nagaosa}, \citenamefont {Felser},\ and\ \citenamefont {Yan}}]{Zhang2019}%
  \BibitemOpen
  \bibfield  {author} {\bibinfo {author} {\bibfnamefont {Y.}~\bibnamefont {Zhang}}, \bibinfo {author} {\bibfnamefont {T.}~\bibnamefont {Holder}}, \bibinfo {author} {\bibfnamefont {H.}~\bibnamefont {Ishizuka}}, \bibinfo {author} {\bibfnamefont {F.}~\bibnamefont {de~Juan}}, \bibinfo {author} {\bibfnamefont {N.}~\bibnamefont {Nagaosa}}, \bibinfo {author} {\bibfnamefont {C.}~\bibnamefont {Felser}},\ and\ \bibinfo {author} {\bibfnamefont {B.}~\bibnamefont {Yan}},\ }\bibfield  {title} {\bibinfo {title} {Switchable magnetic bulk photovoltaic effect in the two-dimensional magnet cri3},\ }\href {https://doi.org/10.1038/s41467-019-11832-3} {\bibfield  {journal} {\bibinfo  {journal} {Nature Communications}\ }\textbf {\bibinfo {volume} {10}},\ \bibinfo {pages} {3783} (\bibinfo {year} {2019})}\BibitemShut {NoStop}%
\end{thebibliography}%

\clearpage

\end{document}